\documentclass{aa}  

\usepackage{graphicx}   
\usepackage{amsmath}    
\usepackage{comment}
\usepackage{siunitx}
\usepackage{float}
\usepackage{txfonts}
\usepackage{hyperref}
\usepackage[nameinlink,capitalise]{cleveref}
\crefname{section}{Sect.}{Sects.}
\def\subinrm#1{\sb{\mathrm{#1}}}
{\catcode`\_=13 \global\let_=\subinrm}
\mathcode`_="8000
\def\upsubscripts{\catcode`\_=12 } 
\begin{document} 

\upsubscripts

   \title{High-cadence monitoring of the emission properties of magnetar XTE J1810-197 with the Stockert radio telescope}
    \titlerunning{High-cadence monitoring of magnetar XTE J1810-197}

   \author{Marlon L. Bause\inst{1},
   Wolfgang Herrmann\inst{2}
          \and
          Laura G. Spitler\inst{1}
          }

   \institute{Max Planck Institut für Radioastronomie,
   Auf dem H\"ugel 69, 53121 Bonn, Germany\\
              \email{mbause@mpifr-bonn.mpg.de}
         \and
             Astropeiler Stockert e.V., Astropeiler Stockert 2-4, 53902 Bad Münstereifel, Germany
             }
    \authorrunning{M. L. Bause et al.}
   \date{Received 07 December, 2023; accepted 12 March, 2024}

 
  \abstract
   {Since the detection of a burst resembling a fast radio burst (FRB) from the Galactic magnetar SGR 1935+2154, magnetars
have joined the set of favourable candidates for FRB progenitors.
However, the emission mechanism of magnetars remains poorly understood.}
   {Observations of magnetars with a high cadence over extended timescales have allowed for their emission properties to be determined, in particular, their temporal variations.
In this work, we present the results of the long-term monitoring campaign of the magnetar XTE J1810-197 since its second observed active phase from December 2018 until November 2021, with the Stockert \SI{25}{m} radio telescope.}
   {We present a singlepulse search method, improving on commonly used neural network classifiers thanks to the filtering of radio frequency interference based on its spectral variance and the magnetar's rotation.}
   {With this approach, we were able to lower the signal to noise ratio (S/N) detection threshold from 8 to 5. This allowed us to find over 115,000 spiky single pulses -- compared to 56,000 from the neutral network approach.
Here, we present the temporal variation of the overall profile and single pulses.
Two distinct phases of different single pulse activity can be identified: phase 1 from December 2018 to mid-2019, with a few single pulses per hour, and phase 2 from September 2020 with hundreds of single pulses per hour (with a comparable average flux density).
We find that the single pulse properties and folded profile in phase 2 exhibit a change around mid-March 2021.
Before this date, the folded profile consists of a single peak and single pulses, with fluences of up to \SI{1000}{Jyms} and a single-peaked width distribution at around \SI{10}{ms}.
After mid-March 2021, the profile consists of a two peaks and the single pulse population shows a bimodal width distribution with a second peak at \SI{1}{ms} and fluences of up to \SI{500}{Jyms}.
We also present asymmetries in the phase-resolved single pulse width distributions beginning to appear in 2020, where the pulses arriving earlier in the rotational phase appear wider than those appearing later.
This asymmetry persists despite the temporal evolution of the other single pulse and emission properties.

}
   {
We argue that a drift in the emission region in the magnetosphere  may explain this observed behaviour. 
Additionally, we find that the fluence of the detected single pulses depends on the rotational phase and the highest fluence is found in the centre of the peaks in the profile.
While the majority of the emission can be linked to the detected single pulses, we cannot exclude another weak mode of emission.
In contrast to the pulses from SGR 1935+2154, we have not found any spectral feature or bursts with energies in the order of magnitude of an FRB during our observational campaign. Therefore, the question of whether this magnetar is capable of emitting such highly energetic bursts remains open.}


   \maketitle
%

\section{Introduction}

Magnetars are highly magnetised neutron stars with magnetic field strengths of the order of \SIrange{e13}{e15}{G, }  which were first introduced by \citet{duncan1992}.
The emission is typically in the X-ray and gamma part of the electromagnetic spectrum, showing transient bursts with rotation periods in the range of approximately \SIrange{1}{12}{s}. Out of the 30 known magnetars in the magnetar catalogue from \citet{olausen2014}\footnote{\url{https://www.physics.mcgill.ca/~pulsar/magnetar/main.html}}, only six have had their radio emission detected.
\citet{rea2012} argued that the detection of radio emission weakens the separation between radio pulsars powered by rotation and magnetars powered by their magnetic energy as the ability for emitting radio emission results of specific conditions of the magnetar identified by the so-called fundamental plane.
The first magnetar that was found to be active in the radio regime was XTE J1810-197: one year after its discovery in 2003 in X-ray by \citet{ibrahim2004}, \citet{halpern2005} discovered radio emission from XTE J1810-197.
Additionally, \citet{camilo2006} found that this radio emission is transient emission and consists of bright, narrow (a few ms in duration) and highly linear polarised single pulses that produced the overall folded profile.
As reported by \citet{camilo2016}, the radio emission disappeared in late 2008 and the first re-detection of radio emission happened ten years later, as reported by \citet{lyne2018}.
This re-appearance of radio emission in December 2018 marked the start of the observational campaign of XTE J1810-197 with the Stockert telescope presented in this article.

The Stockert telescope is a \SI{25}{m} instrument built in 1956. After a hiatus between 1995 and 2005, the instrument was refurbished and equipped with up to date instrumentation. It is now operated and maintained by a group of volunteers, the Astropeiler Stockert e.V. The SEFD of the Stockert telescope was \SI{1000}{Jy} for the observations reported here. After in upgrade in early 2022, the SEFD was improved to \SI{380}{Jy}.

Fast radio bursts (FRBs) are short (micro- to milliseconds in duration) radio pulses of an eceiown extragalactic origin. Since the first detection of an FRB by \citet{lorimer2007}, many FRBs have been detected by several instruments and collaborations.
Some of the FRBs are repeating, which means there are multiple bursts which are coming from the same location in the sky as shown by \citet{spitler2012}. These are the so-called 'repeaters'.
Observable properties of FRBs include their fluence, duration and morphology.
Especially repeating FRBs are typically band limited signals, hence the bandwidth is another observable property.
The morphology of the bursts spans a wide range from simple Gaussian-like bursts to complex multi-component bursts that may show drifts in frequency as for example shown in \citet{pleunis2021} and \citet{hessels2019}.
However, the origin of FRBs remains unknown.
Out of the several theories that try to explain the FRB phenomenon many include magnetars.
An overview of the different progenitor theories can be found in \texttt{frbtheorycat}\footnote{\url{https://frbtheorycat.org}} created by \citet{platts2019}.
Since the detection of an FRB-like burst from the galactic magnetar SGR 1935+2154 by \citet{chimesgr19352020} and \citet{bochenek2020}, magnetars have become a favoured progenitor for at least some of the FRBs. 
On the other hand, the emission mechanism of magnetars is not well understood either. Hence, we study the radio emission of a galactic magnetar with a focus on its single pulses motivated by the possibility of magnetars being potential progenitors of FRBs.

This article is structured as follows.
In \cref{sec:observation}, we describe the data set and the data reductions techniques used in this work, \cref{sec:time_evo_properties} shows our analysis of the data focused on the time evolution and the single pulse properties, \cref{sec:discussion} puts the results of this work in the context of magnetars and FRBs, and \cref{sec:conclusion} gives a summary of the work of this paper.

\section{Observational data}\label{sec:observation}
\subsection{Observations}
\begin{figure}
       \centering
       \resizebox{\hsize}{!}{\includegraphics{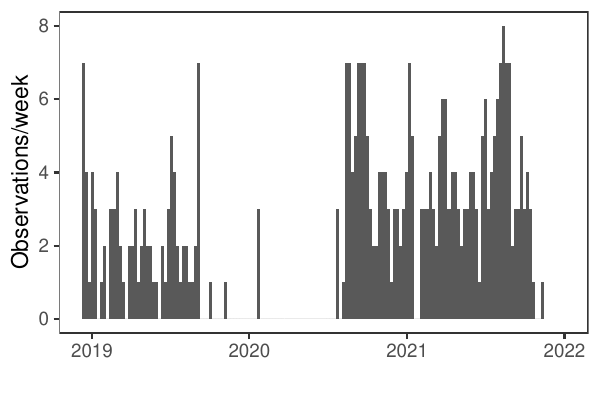}}
       \caption{Number of observations per week over the observational campaign. The observations were stopped between mid 2019 and mid 2020 as the magnetar was not seen in the folded profile.}
       \label{fig:obs_overview}
   \end{figure}

The observations used in this work were performed between 12 December 2018 and 14 November 2021. In total, the data set consists of 347 observations over 339 days, which results in about \SI{1015}{h} of on source time. The duration of a single observation is typically \SI{3}{h}. 
\Cref{fig:obs_overview} gives an overview of the distributions of the observations in the observational campaign.

The cadence of the observations varies over the observational campaign depending on the telescope and staff (volunteers only) availability.
Typically there were two to three observations per week in 2018-19 (phase~1) and three to four observations peer week in 2020-21 (phase~2).
However, there are also weeks with daily observations as well as weeks without observations, for example due to maintenance of the telescope. 
Between mid 2019 and July 2020, there was a gap where only a few observations were done to test whether XTE J1810-197 was visible.
Since the magnetar was not detectable with the ephemerides used at the time of the observations, the observations were stopped until \citet{ATel13840} reported that XTE J1810 was showing a strong radio signal again.
   
The data was recorded in the frequency range \SIrange{1332.5}{1430.5}{MHz}, recording the total intensity 32-bit data with a pulsar fast Fourier transform (PFFTS) backend \citep{barr2013}.
The data were initially stored as 'PFFTS' files, which is the instrument’s specific format.
These were subsequently converted to the filterbank format using the tool filterbank which is part of the SIGPROC package from \citet{lorimer2011}. The resulting data had time and frequency resolution of \SI{218.45}{\micro s} and \SI{586}{kHz} and were stored as 32-bit floats.
Despite the data being recorded with two feeds, the data is averaged to the total intensity during the correlation process due to constraints from the current back-end system.

\subsection{Data reduction}

\subsubsection{Calibration} \label{sec:calibration}
For the majority of the observations since September 2020, a transit scan of the radio galaxy 3C353 at the time of observation is available.
3C353 is a nearby radio galaxy and is expected to have a very stable flux, which makes it a suitable source for calibration purposes.
Its flux density at \SI{1.42}{GHz} is \SI{56.7}{Jy}, as reported by \citet{baars1977}, which translates into a source temperature of \SI{5.67}{K} for the Stockert telescope.
Using the flux density before and after 3C353 is in the beam (OFF) in machine units and the flux density when 3C353 is fully in the beam (ON), the flux calibration factor, $K,$ and the system temperature, $T_{sys}$, can be estimated for flux calibrating the observed data.
For the observations before September 2020, the mean of $T_{sys}$ and $K$ from all observations before September 2020 were used since only an unreliable calibration sources were part of the observation routine.
In any cases where the observation of 3C353 was heavily affected by radio frequency interference (RFI), values from the neighbouring observations were used.

\subsubsection{RFI mitigation}
To mitigate RFI in the data, the \texttt{rfifind} tool, which is part of the PRESTO software package\footnote{\url{https://github.com/scottransom/presto}}, was used.
It identifies intervals potentially containing RFI signals by a statistical analysis.
The following parameters have been used in this process: the time block for each frequency channel is \SI{20}{s}, the threshold for both the rejection and clipping for time domain is $\SI{10}{\sigma}$, the cutoff for the rejection in the frequency domain is $\SI{4}{\sigma}$, a frequency channel is masked entirely if \SI{30}{\%} or more of the time blocks are masked and a time interval is masked if at least \SI{70}{\%} of the frequency channels are masked.

\subsubsection{Search for single pulses} \label{sec:sp-pipeline}
In this work, a single pulse is defined as the spiky, ms-duration emission of the magnetar as it is customary in the FRB-field.
It is important to note that this differs from the definition from a pulsar perspective where typically the emission of an entire rotation referred to as a 'single pulse'.
Since we are interested in the link of magnetar emission to FRBs, we stick to the FRB definition of a single pulse.
To search for single pulses, the filterbank is de-dispersed to a single time series using a dispersion measure (DM) of \SI{178}{pc cm^{-3}} using \texttt{prepsubband} and then searched with \texttt{single\_pulse\_search.py}.
All available default pulse widths (1 to 300 bins equal to \SIrange{0.218}{65}{ms}), no bad block detection and a minimal S/N of 5 were used for the matched filtering in the single pulse search.
Since we used to the default clipping threshold of \texttt{prepsubband}, the ability to recover a MJy pulse (FRB-like) is hindered significantly and the pipeline is fine-tuned to find the fainter single pulses.
Hence, MJy pulses are still detectable but reported with a lower flux density.
A different pipeline design would allow us to see a MJy pulse, as the receiver of Stockert would not be saturated from such a pulse.

The resulting list of pulse candidates contains real single pulses next to RFI and noise events.
A robust filtering method to return those single pulses that are real is needed to further analyse the single pulses. A common classifier in the FRB community is
FETCH\footnote{\url{https://github.com/devanshkv/fetch/tree/master}}, developed by \citet{agarwal2020}, which is   trained with both FRB pulses and pulsar single pulses.
This tool is a machine learning classifier using the dynamic spectrum as well as the DM versus time space of each single pulse candidate to classify single pulses and non single pulses with eleven available models.
We estimate the quality of the different models for our data by creating a set of single pulses from ten observations and manually classifying these single pulses based on their dynamic spectrum and time series via manual inspection.
Based on this set of single pulses, we compare our classification to those of FETCH using the default threshold (\SI{50}{\%}) and all available models (a - k).
In machine learning, the metric to measure the quality of a model consists among others of the precision (fraction of real single pulses labelled correctly by the model), the recall (fraction of real single pulses that has been found by the model), and the FScore (the mean of precision and recall).
The FScore is at best \SI{93.8}{\%} for a single model.
To improve the classifications, we combined the six best models based on their Fscore in groups of three models and required a single pulse candidate to have been labelled as single pulse by at least two models to count as a real single pulse.
The combination of  models a, c, and i gave the best metrics (Fscore = \SI{96.0}{\%}); hence, this was used for the classification step.
A further improvement of the classifications would require overcoming the limitations of the training set that was used for training the models, namely, the signal to noise ratio (S/N > 8) or pulse width (<= 32 time bins).
While the latter can accounted for by applying down-sampling, re-training the models for faint single pulse candidates is difficult since even the manual classification is difficult for low S/N single pulses.

\begin{figure}
    \centering
    \resizebox{\hsize}{!}{\includegraphics{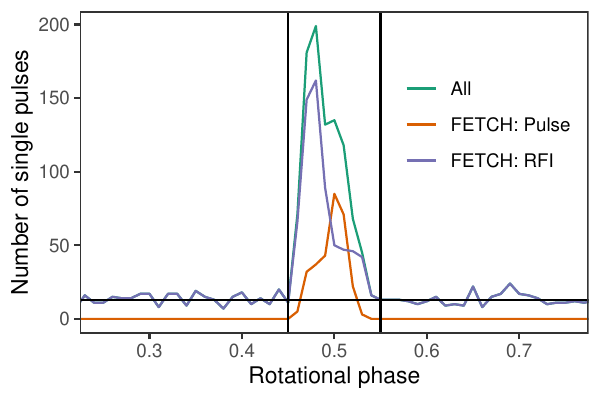}}
    \caption{Distribution of the single pulse candidates in the rotational phase in an observation in April 2021. The three profiles correspond to the labels from FETCH and all single pulse candidates.
    The horizontal lines represents the average number of pulses in the off-pulse window while the two vertical lines mark the emission window of single pulses.}
    \label{fig:sp_filtering}
\end{figure}

As seen in \cref{fig:sp_filtering}, which shows the distribution of single pulse candidates against the rotational phase of one observation, these limitations can lead to a significant under-detection of real single pulse candidates that are found in the search process.
There is a clear peak of pulse candidates the FETCH labelled as RFI or noise at the same phase window where the real single pulses are.
This peak consists of faint single pulses that neither FETCH nor our manual classification is sensitive to -- even though these real pulses actually come from the source.

\subsubsection{Filtering of single pulse candidates}\label{sec:filtering}
To overcome the limitations of FETCH, we developed a filtering technique that makes use of prior knowledge of the single pulse properties as well as the RFI pulse properties.
From the inspection of the FETCH classifications and the phase distribution of single pulse candidates, we find that all detected single pulses from XTE J1810-197 are found in a narrow window of the rotational phase, where no inter-pulses have been detected.
This phase window (on-pulse window) coincides with the window of the pulse profile of the folded time series.
The RFI and noise pulses on the other hand are independent from the rotational phase of the magnetar and thus form a flat baseline that is the average number of RFI and noise pulses in a given phase bin (the horizontal line in \cref{fig:sp_filtering}) is constant (within some fluctuations).
If this baseline goes towards zero, all pulse candidates in the on-pulse region (indicated by the two vertical lines in \cref{fig:sp_filtering}) could be accepted as real single pulse with a low contamination fraction (number of false positives or number of accepted pulses).
We used the contamination fraction as metric to measure the quality of the filtering process.
The number of false positives is estimated from the mean number of pulse candidates per bin in the off-pulse region times the number of bins in the on-pulse region.
The number of accepted pulses is simply the number of pulse candidates in the on-pulse window.
By subtracting the baseline contribution, the number of real pulses can be estimated.
To estimate the on-pulse region, we used the region around the peak of the single pulse histogram that is at least one $\sigma$ higher than the baseline.

To reduce the baseline, we made use of the spectral variance of the pulse candidates and their width. Magnetar pulses are typically broad band, especially in our limited bandwidth of about \SI{100}{MHz}, while RFI pulses are typically narrow band.
This can either be intrinsic to the RFI event itself  for example only a few frequency channels wide or an effect of the de-dispersion, which introduces a sweep in a broadband pulse.
We calculated the frequency variance for each pulse candidate in form of the modulation index $m$ as presented in \citet{spitler2012}.
Following the authors, the modulation index for a broad band S/N = 5 pulse $m \approx$ 3 for the data in this article.
To identify the empirical threshold between broadband and narrowband signals, we inspected the modulation index of single pulse candidates in the on-pulse window, that is, the phase where the real single pulses are.
The vast majority of these single pulse candidates have $m \lessapprox 2.75$, which is particularly the case for those single pulses identified by FETCH.
In contrast, those outside the on-pulse window are distributed towards significantly higher $m$.
Thus, we require $m \leq 2.75$ for a real single pulse, which fits well with the predicted value.

This method does not work if the RFI pulse is much wider than the width that was reported for the pulse.
In this case, the DM-sweep might only cover a part of the pulse candidate and, thus, the modulation index is below the threshold.
To reject these events, we inspect the phase histogram as seen in \cref{fig:sp_filtering} for each individual boxcar width used by \texttt{single_pulse_search.py} and select the widest boxcar width ($w_{max}$) for each observation at which a peak in the on-pulse window is still visible. $w_{max}$ is between \SIrange{15}{33}{ms}.
On the other end of the width spectrum, we can reject those widths which are heavily affected by DM smearing \citep[See][]{pulsarhandbook}, which is about 1.5 time bins and which implies that pulses with a width of 2 or 1 time bins are smeared out and thus physically impossible.
A two bin pulse is physically possible but is still smeared, which decreased the S/N and, thus, the ability to detect such events.
Hence, we rejected all single pulse candidates that have $m > 2.75$, width < 3 time bins (\SI{0.65}{ms}) and width > $w_{max}$.
All remaining single pulses in the on-pulse window are considered as real single pulses.

\begin{figure}
    \centering
    \resizebox{\hsize}{!}{\includegraphics{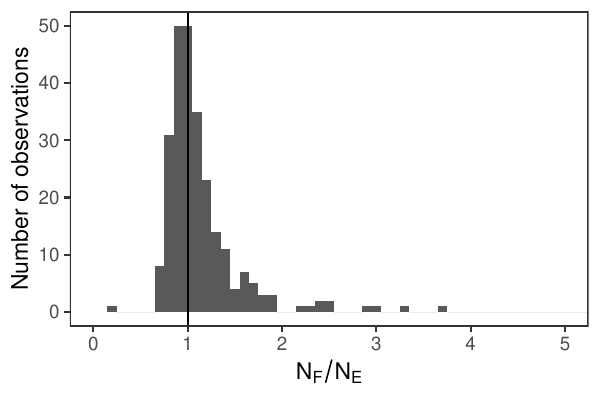}}
    \caption{Histogram of the fraction of measured false positive pulses ($N_F$) and the expected false positives ($N_E$) for each observation with detected single pulses.}
    \label{fig:filtering_fp_noise}
\end{figure}

After the filtering as described above, \SI{85}{\%} of the observations had a contamination of less then \SI{10}{\%}, and \SI{66}{\%} have a contamination fraction of less than \SI{5}{\%}.
To see whether the remaining false positives are noise pulses (which are not targeted by the filtering method described above), we compared the number of false positives with the number of noise events expected from our pipeline and observations in the on-pulse window.
We used the 5 $\sigma$ probability assuming Gaussian noise and multiply this with the number of time series bins of the specific observation, the number of widths used for that observations, and the duty cycle fraction as we are only accepting events within the on-pulse region.
In the case of a \SI{3}{h} observation, which is the average duration of our observations, the estimated number of false positives is about 20 pulses, depending on the number of widths rejected in the filtering process.
\Cref{fig:filtering_fp_noise}, shows the distribution of the false positives measured from the filtering method ($N_F$) and those estimated assuming Gaussian noise ($N_E$) for all observations, where the respective number of widths and duration was used.
The majority of the observations lay around the expected number of false positives resulting from the noise distribution (close to unity).
This suggests that the remaining false positives and thus the contamination are predominantly caused by noise events for these observations.
We expected noise pulses to follow a distribution in their 'pulse properties', whereas RFI typically has distinct fluences and pulse widths and would thus influence the single pulse populations by producing artificial peaks.
However, there are also observations where the number of measured false positives is more than double the estimated number of false positives.
These are partially short observations (observing time << \SI{3}{h}) that are dominated by small number statistics, as well as observations that have been heavily affected by a radar system.
Three observations are so heavily affected by the radar system that the contamination fraction is higher than \SI{15}{\%} and we rejected these three observations.

The majority of the further analysis is focused on the single pulse population properties.  To remove the noise pulse contribution to the distributions (and thus obtain a less biased view of them), we estimated the properties of the noise pulses or generally those of the remaining baseline by looking at the off pulse region.
By making histograms with the same binning as done for the single pulses properties, we can correct those of the single pulse properties by re-scaling the baseline histograms from the off-pulse region to the on-pulse region.
Thus, we can simply subtract the scaled baseline histogram from the single pulse property histogram and thus receive an (ideally) uncontaminated distribution of the single pulse properties.
This filtering gives about 115,000 (120,000 without noise correction) single pulses for further analysis while FETCH gave only about 56,000 single pulses.
For the rest of this work, only the filtered set of single pulses is considered.

\subsubsection{Estimation of single pulse properties}\label{sec:estimation-sp-properties}
The three main single pulse properties that we consider in this work are the width, the fluence, and the mean flux density.
The width of the single pulse is directly taken from the boxcar width reported by PRESTO.
For the fluence of the single pulses, we sum over the de-dispersed time series at the location of the pulse, after subtracting the baseline contribution in the area of the pulse (800 - 1000 bins away from the pulse centre) and multiply the sum by the calibration factor of the observation.
The mean flux density is obtained by the division of fluence of the single pulse by its width.

While the number of single pulses found in our data set is large, it is important to consider that only some of the overall single pulse population is detectable and hence our data set is incomplete.
There are two ways of incompleteness. 
The first gives the fraction of real single pulses that have been found by the pipeline ('recall').
The second follows from the parameter space of the single pulses as some regions of fluence and width fall below our S/N threshold and are thus not detectable.
We focus on the parameter space incompleteness and base our completeness limits on this type of incompleteness. 
The completeness limits arise from the time and frequency resolution as well as from our data reduction technique.
We are directly limited by the time resolution of the data $t_{data}$ that is the most narrow width that we could find is \SI{0.218}{ms}.
More narrow pulses would be smeared to the resolution at a cost of S/N and are thus hard to detect.
However, in combination with the frequency resolution of the telescope and the dispersion of the signal, the DM smearing \citep[See][]{pulsarhandbook} $t_{DM}$ is about 1.5 time bins (as noted in \cref{sec:filtering}).
As scattering is negligible, the time resolution of the data $t$ follows from $t = \sqrt{t_{data}^2 +t_{DM}^2} \approx 1.8 $ bins.
Therefore, the ability to find pulses of a width of two bins is significantly worse than wider pulses as they are smeared and thus have a lower S/N and partially fall under our detection threshold and pulse widths of 1 bin are physically not possible.
Hence, we set the completeness limit for bursts width to three bins which corresponds to \SI{0.654}{ms}.
We note that this limitation is also used in the filtering process for this reason.

The second kind of limitations follow from our search for single pulses. Our S/N threshold in the searching stage is 5 and hence all single pulses with a S/N below 5 are systematically not detected.
We can transform the S/N into physically meaningful and telescope independent values using the radiometer equation.
To estimate the minimum fluence, $F,$ or average flux density of the single pulse from the minimal S/N threshold ($S/N_{min}$), respectively, we use:
\begin{equation}
    \centering
    \label{eq:radiometer}
    F = \frac{(S/N) T_{sys}}{G \sqrt{n_{pol} B}} \sqrt{\delta t},
\end{equation}
where $G$ is the gain of the telescope, $T_{sys}$ is the system temperature of the telescope, $n_{pol}$ is the number of recorded polarisations, $\delta t$ is the pulse width, and $B$ is the width of the bandpass.
\Cref{eq:radiometer} shows that, the completeness threshold for the fluence is a function of pulse width and grows $\propto \sqrt{\delta t}$ that is the overall fluence completeness is given by the largest significant pulse width (\SIrange{10}{15}{ms}).
For the average flux density, the completeness limit follows from \cref{eq:radiometer} by dividing $F$ by the pulse width, $\delta t$, which is the limiting flux density that scales with $\propto t^{-0.5}$.
Hence, the overall mean flux density completeness threshold is given by the most narrow significant pulse width (\SI{0.654}{ms}).
This gives overall completeness limits of about \SIrange{50}{80}{Jyms} and about \SI{20}{Jy,} respectively.
Rather then specifying a single completeness limit, we specify when specific pulse widths are below the completeness threshold for the fluence and mean flux density distributions.

\subsubsection{Folding the time series}\label{sec:folding}
The de-dispersed time series was folded manually by slicing it into chunks of the (topocentric) period obtained from the corresponding ephemerides and adding them together.
Four different sets of ephemerides based on those published by \citet{levin2019} and \citet{caleb2022} and have been provided to the authors.
The first set was used for the observations until MJD 59000, the second set for the observations between MJD 59000 and MJD 59246, the third for the observations after MJD 59246 until MJD 59469 and the fourth for the observations after MJD 59469.
Any rotation that was not fully recorded (e.g.  due to RFI masking or the start and end of the observation) was rejected.
This method also allows us to fold only specific rotations, for example, those with detected single pulses.
In total we calculate three profiles from the time series: one with all (RFI-free) rotations, a second consisting of the rotations with a detected single pulse, and a third profile from rotations without detected single pulses.
Similarly, the emission from the single pulses is folded into a profile. 
In this case, only the baseline removed part of the rotation which includes the single pulse (about a few ms) was used.
From these profiles, the average flux density and the total received fluence can be obtained after calibration and baseline removal (for the folded rotations).

\subsubsection{Profile alignment}\label{sec:profile_alignment}
A profile alignment is necessary to compare the emission at specific rotational phases of the magnetar.
However, magnetars are generally less stable so that the ephemerides are not as precise as for pulsars and the ephemerides are generally only valid for specific time ranges.
Additionally, the uncertainty of the clock in the telescope back-end, produces a random shift of the order of a few milliseconds in phase between individual observations.
Hence, the profile alignment was done manually by inspection the folded profiles.
We made use of the structure of the profiles, which is comparable between consecutive observations, such as the location of the peak(s) in the folded profile. We added a phase shift so that the phase location would match the previous observations and we set the centre of the profile to 0.5 of the rotational period.
In this profile alignment process, we assumed that the emission window in the rotational phase is stable, so that is there is no event with a large glitch within our observational campaign as long as the profile has a similar form.
This is supported by the results from \citet{caleb2022}, who observed the magnetar over a similar time span.
They find that consecutive observations show a dominant main component at a stable position in phase.
The uncertainty of the shifting is about \SI{20}{ms} (about \SI{0.4}{\%} of the rotational period), so that profile features with a duration shorter than this are smeared out.

For the single pulses, we calculate a phase histogram.
This is done by calculating at which rotational phase the central bin of the single pulse lies and then counting the number of single pulses in a phase bin.
The resulting 'profile' of the single pulse histogram shows features in the form of peaks that are analogous to the folded profiles.
Thus, they can be phase aligned by the same procedure as for the folded profiles.
However, as they do not contain noise from the time series, the structure in the individual 'profiles' is clearer and the accuracy is significantly improved.
This allows for an improvement of the uncertainty to about \SI{5}{ms} (\SI{0.1}{\%}) of the rotational period.

\section{Emission properties}\label{sec:time_evo_properties}
This section presents the analysis of the single pulses as well as the folded profiles.
The subsections focus on specific aspects of the emission. To draw conclusions on the emission mechanism, we start the discussion with a section where we merge the individual results into a single picture in \cref{sec:integrationg-the-parts}.
\subsection{Average flux density and the pulse rate}
\begin{figure*}
\centering
\includegraphics[width=17cm]{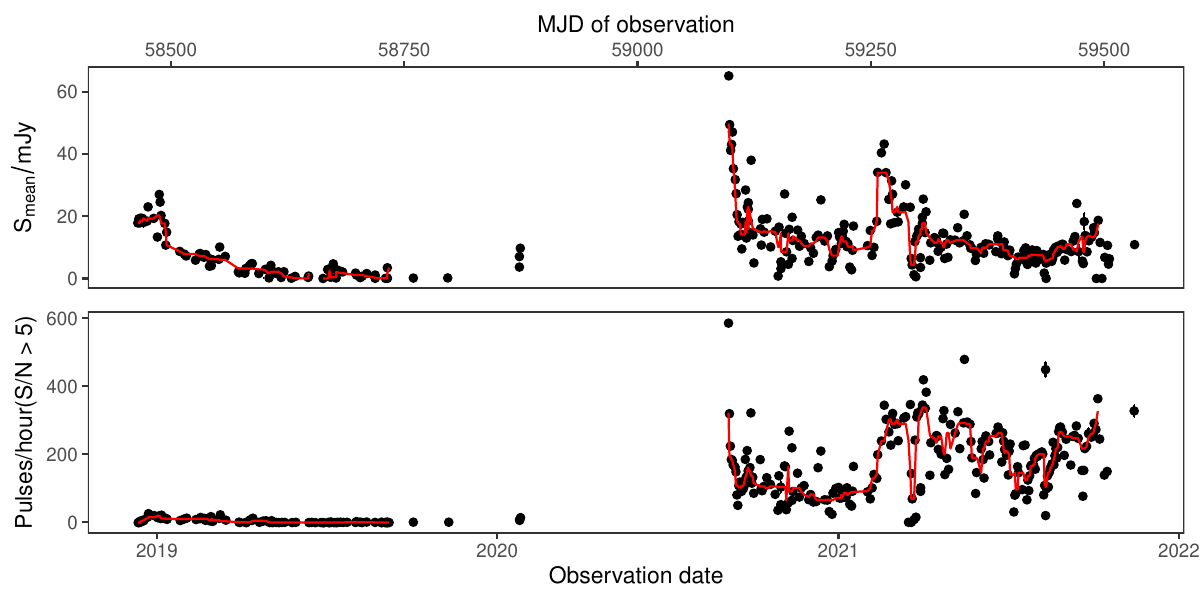}
\caption{Mean flux density, $S_{mean}$, of the folded profile (top) and the number of detected single pulses per hour (bottom) for each observation. The red line shows the corresponding running median with a window of seven observations.}
\label{fig:timeseries}
\end{figure*}

The time evolution of the average flux density ($S_{mean}$) and the pulse rate gives an impression on how the overall radio emission of the magnetar changes in our observational campaign.
The average flux density is calculated from the folded profiles using all rotations not affected by RFI as described in \cref{sec:folding}.
For the pulse rate, we use the approach displayed in \cref{fig:sp_filtering}, namely, we did not apply the filtering; instead, we subtracted the of-pulse baseline counts from each bin in the on-pulse window and sum over them to get the number of pulses in one observation.
\Cref{fig:timeseries} displays the time evolution of the both quantities over the observational campaign with example profiles from the different times.

Based on observational cadence and the activity of the magnetar, we can see two distinct phases: phase 1 in 2018/19 and phase 2 after September 2020.
Phase 1 started with an average flux density of about \SI{20}{mJy} but with time the average flux density decreased to the point where the magnetar was too faint to be detectable around May 2019.
In an average observation of \SI{3}{h}, the upper limit for $S_{mean}$ is \SI{3.5}{mJy} for a non-detection.
The rate of detected single pulse in phase 1 is only a few to ten pulses detected per hour and there is no change in time visible.
There are a three observations in January 2020 which show single pulse activity and an average flux density of up to \SI{10}{mJy}.
However, at the time of observation, the ephemerides used to fold the data were too uncertain and thus the folded profile was missed. Hence, observations were not continued and it was only discovered in hindsight that the magnetar was emitting at this time.

In the beginning of phase 2, $S_{mean}$ was with over \SI{60}{mJy} by far the highest measured in our observation campaign but it decreased to a level of around \SI{10}{mJy} within about a month.
In February 2021, the magnetar re-brightend to about \SI{40}{mJy} before dropping to the level of around \SI{10}{mJy} again.
A similar trend is also visible in the pulse rates in phase 2, it started with extremely high rates of 600 pulses per hour but dropped to a level of around 100 pulses per hour.
The re-brigthening increased the pulse rates to around 300 pulses per hour from which the rate decreased as $S_{mean}$ decreased to a somewhat higher rate of around 200 pulses per hour.

The rate of detected single pulses differs significantly between phase 1 and phase 2 that is in observations that have a comparable $S_{mean}$ in the two phases, there are barely any detected single pulses in phase 1 while there are hundreds of single pulses in phase 2.
In addition to the long-term changes seen for  the two quantities, there are oscillations on a smaller timescale after the re-brightening.
These are most notable in the pulse rates where the rate increases and then decreases again within a month.
The most extreme example of this feature is seen on 17 March 2021, where no single pulses (as well as no profile) were seen within the observation.

\subsection{Morphologies of the single pulses}
\begin{figure}
\centering
\resizebox{\hsize}{!}{\includegraphics{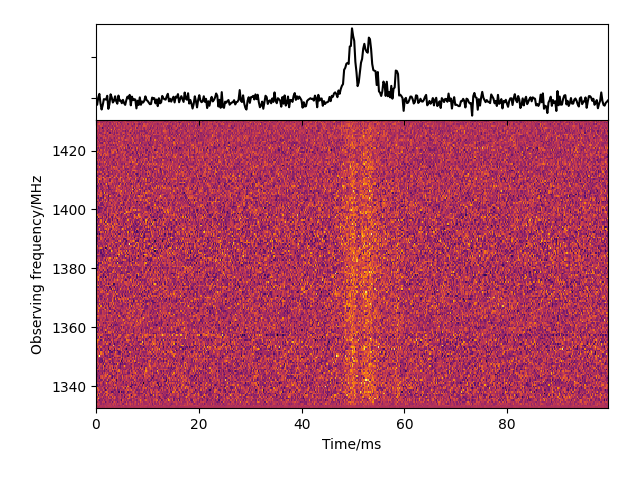}}
\caption{Dynamic spectrum (bottom) and the time series (top) of an example single pulse from the observation on 29 September 2020.}
\label{fig:SP_example}
\end{figure}
The detected single pulses show various morphologies in their dynamic spectra from rather simple 'one-burst' pulses over multiple close bursts as displayed in \cref{fig:SP_example} to more complex multi-component single pulses.
Even within a single observation, the morphology of the detected single pulses varies significantly. This can be seen in \cref{fig:SP_morphologies}, where several single pulses that were observed during the observation on 2021 September 29.
Some of these single pulses have a similar morphology as the complex FRBs seen by \citet{pleunis2021}.
In particular, burst 7 looks similar to C356 in \citet{jahns2023} from FRB20121102A.
However, all single pulses are broadband in the observed band and no  downward drift in frequency of consecutive sub-pulses ('sad trombone effect') is observed (commonly seen in FRBs) has been observed in our data set.
Given the high number of detected single pulses and the variety of morphologies, we limited ourselves to statistical analysis of the single pulse properties.

\subsection{Time evolution of the emission}
The large amount of on source time with high cadence and the high rates of detected single pulses give us the basis to perform statistical analyses of the single pulse properties, which are the (boxcar) width, the fluence, and average flux density, over time.
For this, we defined the data groups by combining the individual sets of single pulses and folded profiles from consecutive observations.
This gives us larger single pulses data sets and high S/N profiles to make robust statistical statements of the time evolution.
When forming these data groups, we kept the on source time of observations comparable between the groups (around \SI{60}{h}), with the exception of the phase 1 observations where we grouped all observations that had a detection together due to the low number of detected single pulses.
Moreover, we made sure that the folded profiles of the observations combined to a data group is comparable.
One exception is data group 5, during which the profile underwent significant changes.
Given a varying density of observations and observation length, the number of observations in a group varies between 18 and 62.
An overview of the time spans, number of observations and single pulses in the groups is given in \Cref{tab:dataranges}.
\begin{table}
\centering
\caption{Data groups used time evolution analysis. For each data group (identified the number in the first column), their time span ('From' and 'To'), the total on source time ('Time'), the number of days over which the observations are distributed ('Days'), the number observations ('Obs.') and number of single pulses ('SPs') for each group is listed.}
\label{tab:dataranges}
\begin{scriptsize}
\begin{tabular}{llllllll}
\hline
 No. & From           & To         & Time (h)  & Days   & Obs. & SPs                     \\
  \hline\hline
 1 & 2018-12-12     & 2019-05-04 & 160     & 144      & 47   &    1199                 \\
 2 & 2020-09-04     & 2020-09-21 & 64.96   & 18       & 16   &   11158                 \\
 3 & 2020-09-22     & 2020-11-22 & 64.84   & 62       & 31   &    9186                 \\
 4 & 2020-11-25     & 2021-02-04 & 66.16   & 42       & 29   &    5725                 \\
 5 & 2021-02-06     & 2021-03-26 & 65.01   & 49       & 23   &   14835                 \\
 6 & 2021-03-27     & 2021-05-01 & 61.10   & 36       & 22   &   15670                 \\
 7 & 2021-05-06     & 2021-06-11 & 59.53   & 37       & 18   &   12822                 \\
 8 & 2021-06-12     & 2021-07-22 & 60.81   & 41       & 22   &   10769                 \\
 9 & 2021-07-23     & 2021-08-13 & 62.77   & 22       & 22   &    9982                 \\
10 & 2021-08-14     & 2021-09-11 & 63.74   & 29       & 21   &   13544                 \\
11 & 2021-09-13     & 2021-11-14 & 52.70   & 62       & 20   &   10425                 \\
\hline
\end{tabular}
\end{scriptsize}
\end{table}

\subsubsection{Mean folded profiles}\label{sec:folded-profiles}
\begin{figure}
\centering
\resizebox{\hsize}{!}{\includegraphics{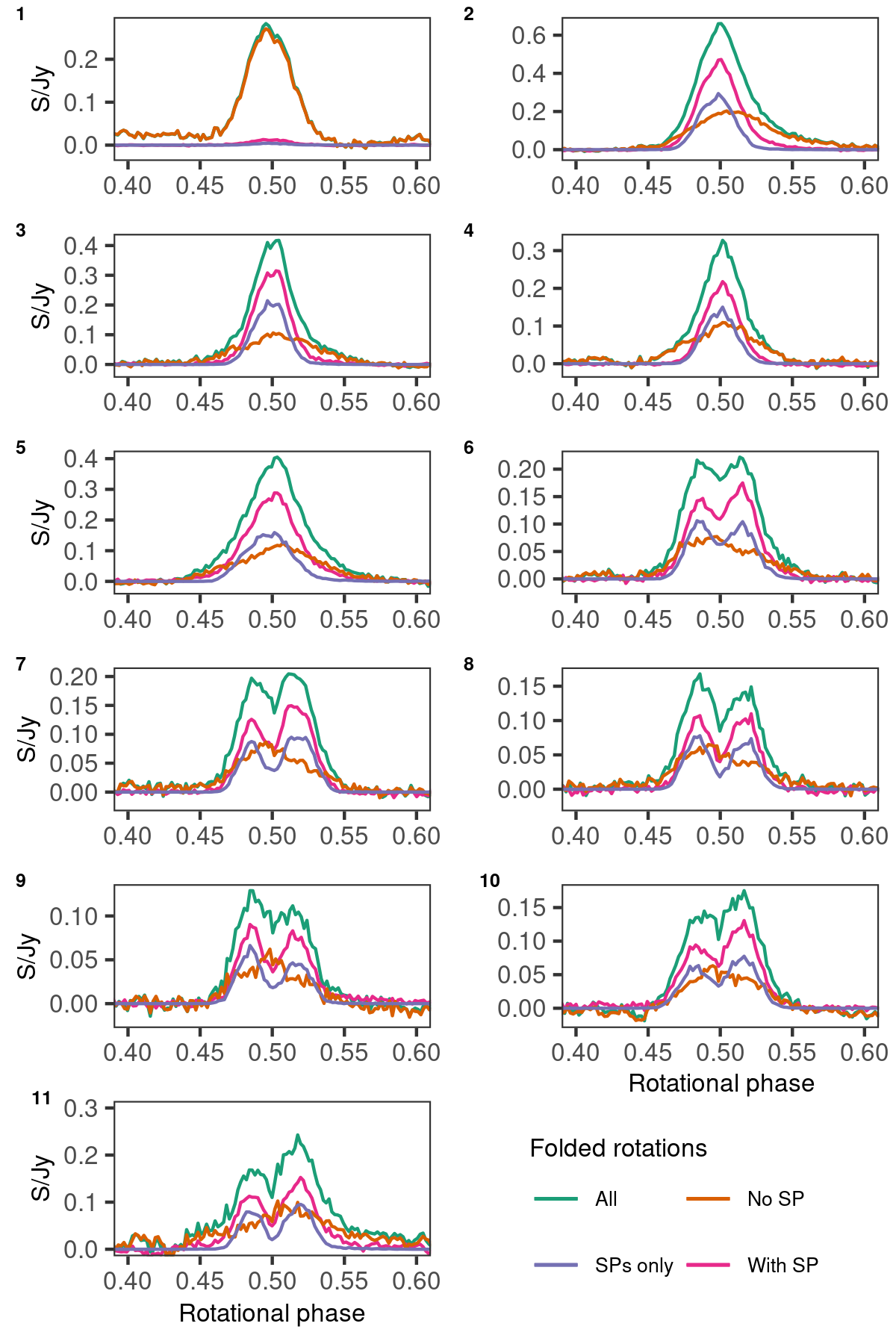}}
\caption{Mean profiles for the observations in the data groups presented in \Cref{tab:dataranges}. For each range, the overall profile resulting from all rotations (All), the rotations without a detected single pulse (No SP) and those with a detected single pulse (With SP) are shown. Additionally, the profiles from the single pulse emission only (SPs only) are included.}
\label{fig:mean_profiles}
\end{figure}

We defined four different profiles for an observation. Three originated from the folding of the time series using all rotations ('All in \cref{fig:mean_profiles}), rotations with a detected single pulses ('With SP' in \cref{fig:mean_profiles}) and rotations without a detected single pulse ('No SP' in \cref{fig:mean_profiles}).
In all cases, rotations affected by RFI or masking were discarded.
We produced the fourth profile ('SPs only') by folding emission of only the single pulses, which is only the time series bins occupied by the detected single pulses.
All four profiles are normalised by the total number of rotations and calibrated.
Hence, adding the profiles from the rotations with a detected single pulse and those without will add to the overall profile.
The profiles of the data groups are phase aligned by the method described in \cref{sec:profile_alignment} with respect to each other.
It is important to note that these profiles are not equal to profiles that are formed using a phase connected timing solution.
The emission region might drift in phase between observations, which would not be possible to recover from our alignment process.
The timing analysis in \citet{caleb2022} overlaps with our observations from the start in December 2018 until November 2020, during which they see three abrupt changes in phase of the dominant profile component.
Our approach would align these, as they do not coincide with significant changes the in folded profiles.
In our case, the alignment is necessary for obtaining a statistically significant sample for the analysis of the phase dependence of the single pulse properties.
Nevertheless, we cannot exclude a drift of the main component in phase in between the observations.
In this work, we assume that such a potential phase drift does not affect the properties of the emission as long as the shape of the four observed profiles remains stable.

The four mean profiles of the magnetar were formed by averaging the calibrated average profiles of the observations in the group for each data group from \Cref{tab:dataranges}.
 We note that the profile for data group 1 includes only observations prior 15 March 2019 (MJD 58557), as \citet{caleb2022} showed a phase shift of the profile at \SI{1.5}{GHz} relative to  \SI{6}{GHz} at the end of March 2021.
Since the number of single pulses, the profile intensity (see \cref{fig:timeseries}) and the number of observations is low, we decided against grouping the observations after 15 March 2019 into a separate profile.
The resulting mean profiles are displayed in \cref{fig:mean_profiles}.
The profiles in data group 1 (which is equal to phase 1) and data groups 2 to 5 show a single peak.
From data group 6 on, the profile consists of two distinct peaks.
The strength of the peak(s) varies, but the overall shape remains similar in data group 2 to 5 and 6 to 11.
The features of the overall profiles (green) in \cref{fig:mean_profiles}, are generally dominated by features of the profiles of the rotations with a detected single pulses (magenta), which itself follows the features of the profile from the single pulses only (purple).
On the other hand, the rotations without a detected single pulse form broad and rather featureless profiles (indigo), which in case of the data groups 2 to 5 contribute at the rising edge and the trailing edge of the over all folded profile where rotations with a detected single pulses are not contributing significantly to the overall profile.
An exception from this behaviour is data group 1, in which only a few rotations contain a detected single pulse and thus the rotations without a detected single pulse dominate the overall profile.
Hence, we focus on the profiles from the other data groups (from phase 2) observations.

Since the profiles from the single pulses and the rotations with a detected single pulse have similar shapes but different amplitudes, there must be additional emission to what we detected as single pulses.
This emission must have properties capable of producing a similar profile as the detected single pulses.
Given that we know that our single pulse search and classification are only complete in a certain parameter space of the single pulses (as shown in \cref{sec:estimation-sp-properties}) the remaining emission could originate from single pulses with properties we cannot detect; for example, this may include very narrow (width < \SI{0.65}{ms}), intrinsically broad pulses (width > \SI{65}{ms} that is wider than \SI{16}{\%} of the typical on-pulse window) and faint single pulses of the known parameters.
These undetected single pulses occur also in rotations without a detected single pulse and add to a profile.
However, the edges of some overall profiles extend beyond the single rotations profiles and overlap with the profiles from the rotations without a detected single pulse.
One clear example is data group 2, where the rotations without a detected single pulse extend for \SI{10}{\%} of the rotational period after the peak of the profile and requires an additional source of emission.
The analysis in the following sections focuses on understanding the properties of the observed single pulses to see whether all the emission can be linked to the detected single pulses or whether there is a weak and broad additional emission mechanism.

\subsubsection{Fraction of fluence from observed SPs}
\begin{figure}
\centering
\resizebox{\hsize}{!}{\includegraphics{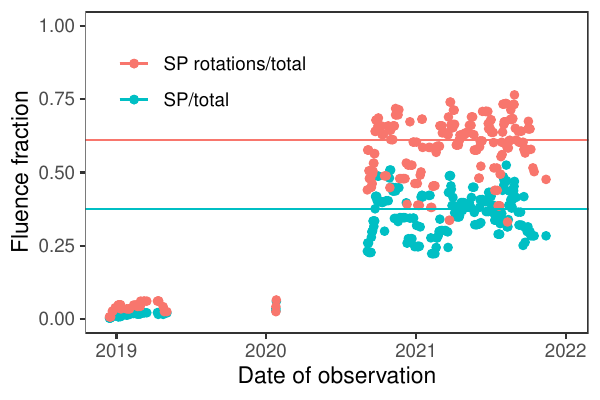}}
\caption{Fraction of the overall received fluence in single pulses (SP/total) and the rotations with a detected single pulse (SP rotations/total) for each observation.}
\label{fig:energy_fraction}
\end{figure}

To quantify the amount of fluence that we can detect as single pulses, we calculated the total fluence of the single pulses, the overall profile, and the profile from the rotations with a detected single pulse.
\Cref{fig:energy_fraction} shows the fraction of the total fluence from the detected single pulses and the fraction of the total fluence from rotations, in which a single pulses was detected.
Observations without a detection of single pulses or a clear folded profile were omitted.
For phase two, these fractions remain roughly constant but there is a high overall scatter, which is a consequence of the high variation of the magnetar emission from observation to observation.
On average, \SI{61}{\%} of the total received fluence is from the rotations that include detected single pulses and about \SI{37}{\%} of the total fluence is from the detected single pulses themselves for the observations in phase 2.
Consequently, \SI{39}{\%} of the received fluence cannot be linked to detected single pulses.
For the phase 1 observations, the fractions are of the order of a few percent and similar to the profiles, the emission is not dominated by detected single pulses

\subsubsection{Single pulse properties}
\begin{figure}
\resizebox{\hsize}{!}{\includegraphics{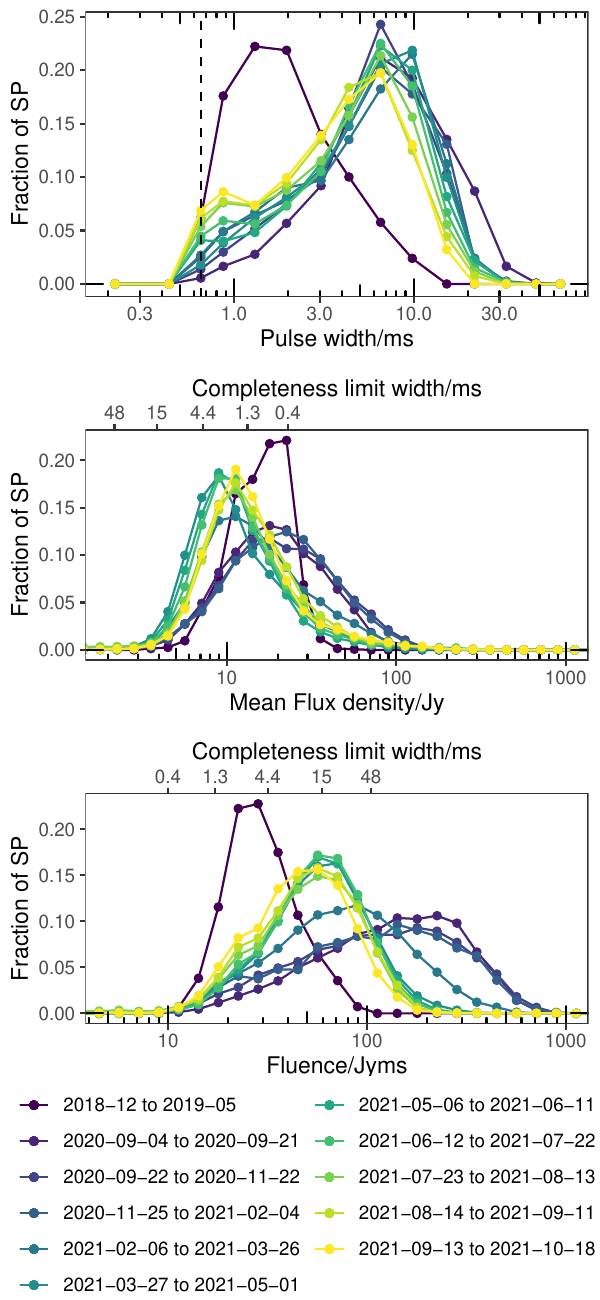}}
\caption{Distribution for the single pulse properties width (top) mean flux density (middle), and fluence (bottom) for the single pulses in each data group as listed in \Cref{tab:dataranges}. For the mean flux density and the fluence, the flux density or the fluence where specific pulse widths with a S/N below the detection threshold are marked.}
\label{fig:SP_property_distributions}
\end{figure}
The single pulse properties of interest are: the pulse width, the fluence, and the mean flux density of the pulse.
In \cref{fig:SP_property_distributions}, we present the distribution of these quantities for each data group defined in \Cref{tab:dataranges}.
The top panel of \cref{fig:SP_property_distributions} shows the pulse width distributions.
We have detected single pulses in the range of \SIrange{0.65}{30}{ms} (the searched window is \SIrange{0.65}{65}{ms}) over our observational campaign.
The absence of pulses below the dotted line is a direct consequence of our limited time resolution due to the DM smearing.
The higher limit at around \SI{20}{ms} may be a consequence of the search identifying a wide pulse as two, narrower sub-pulses.
To test this, the pulses with a waiting time (the time between two consecutive pulses) of less then \SI{20}{ms} were counted.
If all of them would pertain to a single event (i.e. a single pulse wider than \SI{20}{ms}), this would increase the fraction of these wider pulses by about \SI{8}{\%} for the data groups 2 to 5 and \SI{5}{\%} for the data groups 6 to 11.
Hence, only a low fraction of pulses is detected as multiple events rather than one wide pulse and the limit is indeed physical.
This is also well above the limitation of FETCH, which was only trained with up to 32 time bins (\SI{7}{ms}).

The pulse width distributions between the phase 1 and phase 2 single pulses differs significantly: in phase 1, the detected single pulses peak at about \SIrange{1}{3}{ms}, while in phase 2 the distributions peak around \SIrange{5}{10}{ms}.
Within phase 2, there is a clear trend towards more narrow single pulses with time.
The width distribution of data group 2 (September 2020) is almost a mirrored version of the data group 1 width distribution, that is, it has a low fraction (about \SI{20}{\%}) of detected single pulses in the peak region (below \SI{4}{ms}) of data group 1.
This fraction gradually increases over the time and for the data group 11, about \SI{50}{\%} of the single pules are detected with pulse width below \SI{4}{ms}.
The form of the pulse width distribution changes from a single mode distribution with a peak at \SI{6.5}{ms} in data group 2 to a bimodal distribution with peaks at \SI{0.9}{ms} and \SI{6.5}{ms} for the latest data groups.
The peak at lower widths has to be taken with caution since the completeness limit at \SI{0.6}{ms} due to DM smearing is only one bin away.
Hence, it is unclear how the distributions continue beyond this limit.

The middle panel of \cref{fig:SP_property_distributions} shows the mean flux density distributions for the data groups.
The mean flux density of the detected SPs ranges from a few Jy to about \SI{300}{Jy} with some outliers reaching \SI{600}{Jy} in the observations of data groups 5 to 11.
Similar to the width distributions, the phase 1 observations are offset from the phase 2 distributions and cover only a small parameter space of up to \SI{50}{Jy}, which lies almost entirely in the incomplete region.
Hence, we are only sensitive to the very brightest single pulses of this period and there are potentially many more single pulses below our detection threshold.
For the phase 2 single pulses, the data groups split into two clusters: data groups 2 to 4 (profile with a single peak, before the sudden turn-off of emission) and 6 to 11 (profile with two peaks, after the sudden turn off of emission).
The data groups 2 to 4 have a broad mean flux density distribution peaking around \SI{20}{Jy}.
The peak in the mean flux density coincides with the overall completeness limit.
Therefore, the location of the peak may be a consequence of our sensitivity rather than a property of the emission mechanism.
For the data groups 6 to 11, the distributions peak at \SI{10}{Jy}, well in the incomplete region, and are in general fainter than both the data groups 2 to 4 and the phase 1 pulses.
Moreover, there is a slight trend towards brighter single pulses within the groups 6 to 11 with time.
The distribution of data group 5 is a mix of the distributions of the clusters prior and after.
This indicates that the change in the shape and the dimming of the single pulses happened on a time scale of about 1.5 months or even shorter.
For the part of the distribution that is complete, the data cannot be approximated by a simple mathematical distribution for example a power law or a log-normal distribution.

The bottom panel of \cref{fig:SP_property_distributions} shows the distributions of the single pulse fluence, which is the product of the mean flux density and the pulse width, for each data group.
Generally, we find single pulses with fluences in the range of \SIrange{10}{1000}{Jy\,ms}.
The distributions are showing the same clusters as seen for the mean flux density: the data groups 2 to 4 (profile with a single peak, before the sudden turn off of emission) and 6 to 11 (profile with two peaks, after the sudden turn off of emission), while data group 5 is a mix of these two.
The phase 1 observations are off from the other observations with fluences below \SI{100}{Jy\,ms}.
Similar to the mean flux density, the phase 1 and the data groups 6 to 11 distributions lay entirely or to a large fraction in the incomplete region.
The distributions for the data groups 2 to 5 show a plateau at about \SIrange{100}{300}{Jy\,ms} and falls of towards higher fluences as well as lower fluences since the incompleteness sets in towards lower fluences.
The shapes of the fluence distributions resemble the shape of the width distributions but skewed or stretched by the mean flux density that is broadened for the data groups 2 to 5 but skewed for the data groups 1 and 5 to 11.
Also, the trend of less high energetic pulses and more low energetic pulses with time within the data groups 6 to 11 follows the evolution of the width distributions and outruns the trend within the mean flux density distributions.
As for the mean flux density, the distribution cannot be described by a simple mathematical distribution such as a power law or a log-normal distribution.

\subsection{Phase dependence of single pulse properties}
The  analysis described in this work is focused on the single pulse population as a whole. 
However (as discussed in \cref{sec:folded-profiles}), single pulse emission is lacking at specific rotational phases.
This could be caused by a phase dependence of specific single pulse properties.
In this section, we investigate the phase dependence of the single pulse properties for the same sets presented in \Cref{tab:dataranges}.

\subsubsection{Width-resolved single pulse phase histograms}\label{sec:clustering_SPs_rotation}
\begin{figure}
\centering
\resizebox{\hsize}{!}{\includegraphics{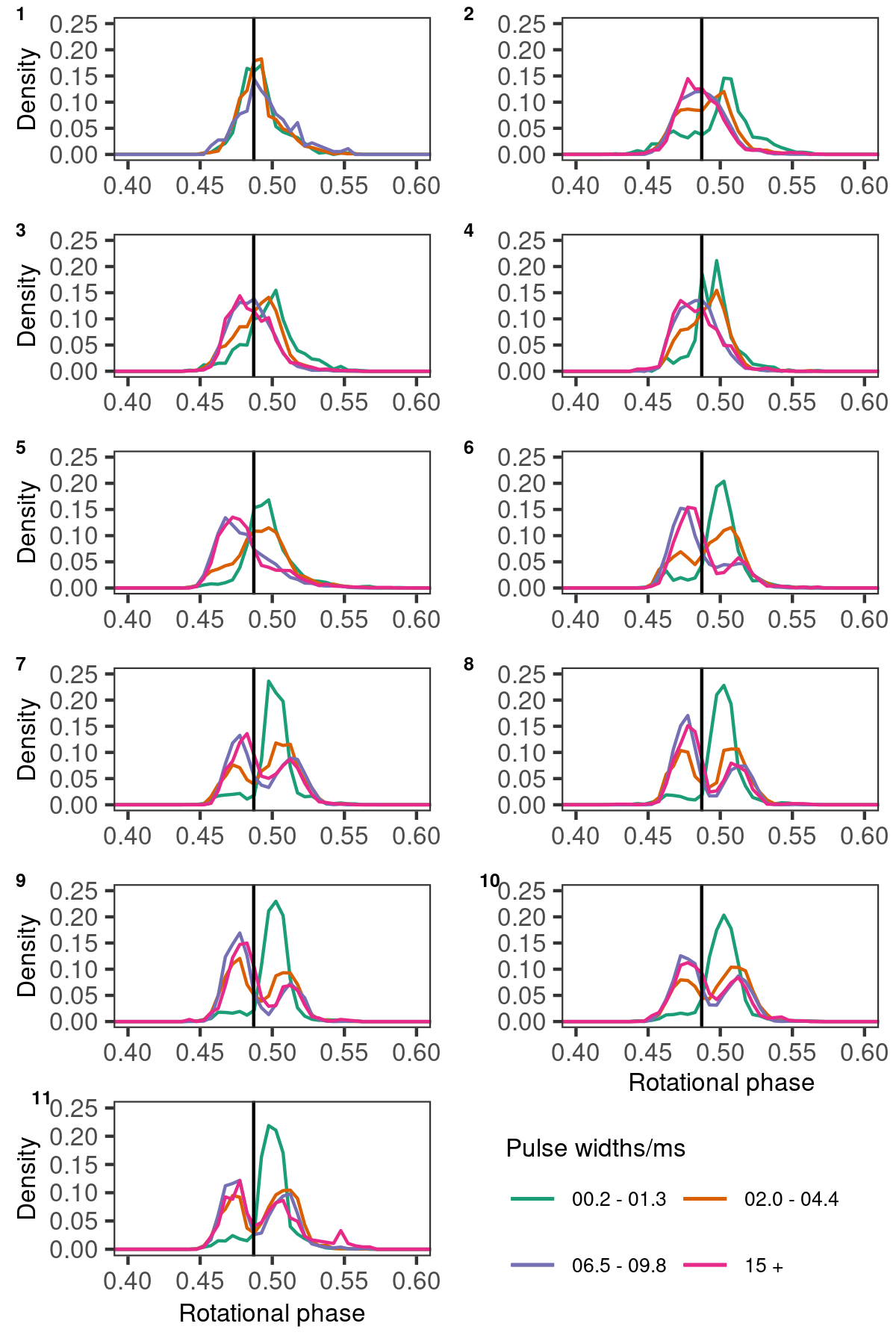}}
\caption{Phase histograms of the rotational phase where the single pulses of the different width ranges have been detected for the data groups listed in \Cref{tab:dataranges}. The vertical line represents where the single pulse samples where split for the analysis in \cref{sec:halves}.}
\label{fig:SP_width_resolved_profiles}
\end{figure}

We begin our investigation of the phase dependence of single pulse properties by looking at the widths of the detected single pulses.
For this, the phase of the central bin of the single pulse candidate is calculated and histograms of the rotational phase for four different pulse width ranges (< \SI{1.3}{ms}, \SIrange{2}{4.4}{ms}, \SIrange{6.5}{9.8}{ms}, and > \SI{15}{ms}) were created for each data group.
\Cref{fig:SP_width_resolved_profiles} shows these phase histograms, where each histogram is normalised by the total number of single pulses in the respective width range.
This emphasises where in the rotational phase each width range is most active independent of the fraction of pulses from the respective group.
Overall, the histograms agree with the respective folded profiles from \cref{fig:mean_profiles}.
However, the phase histograms show significant asymmetries for the single pulse widths per phase for the observations in phase 2, but not for the phase 1 observations, where the histograms are aligned.
The profiles from phase 2 have in common that the narrowest single pulses are detected at rotational phases offset from the wider pulses.
In the observations of the data groups 2 to 5 (September 2020 until March 2021), the narrow single pulses are lagging behind the wider single pulses.
In the data groups 6 to 11, they occur almost in between the two peaks in the observations 
The changeover coincides with the sudden turn off around 17 March 2021, which is visible in \cref{fig:timeseries}, and the change in the single pulse properties in \cref{fig:SP_property_distributions}.
Comparing all width ranges for the data groups 2 to 5, there is a trend seen for wider pulses at earlier phases to narrower pulses at later phases.
Considering the two peaks separately in the data groups 6 to 11, the trend is inverse, that is the narrow single pulses occur towards the beginning of the respective peak, while the more wider pulse occur slightly later within the rotational phase.
However, there are only few single pulses of the most narrow width range detected in the first peak.

\subsubsection{Single pulse properties in the distribution split}\label{sec:halves}
\begin{figure}
\centering
\resizebox{\hsize}{!}{\includegraphics{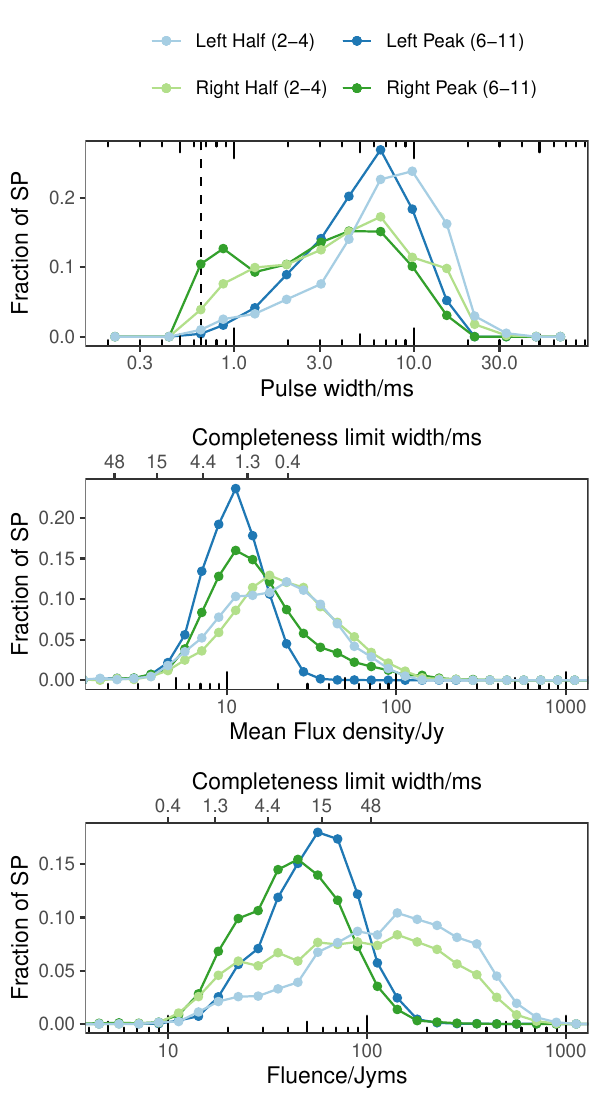}}
\caption{Distribution for the single pulse properties width (top) mean flux density (middle), and fluence (bottom) for the single pulses in the left half and the right half of the data groups 2 to 4, which show a single peak in the folded profile, and the data groups 6 to 11, which show two peaks in the folded profile.}
\label{fig:SP_props_peaks}
\end{figure}

Motivated by the asymmetric width distribution  in \cref{fig:SP_width_resolved_profiles}, we split the single pulses in each data group into two groups at the corresponding line in \cref{fig:SP_width_resolved_profiles} so that we have the single pulses detected in the left half and the right half of the rotational phase.
We then combine these two sets of single pulses from the halves of the data groups where the folded profile shows only one peak (data groups 2 to 4) and where the folded profile shows two peaks (data groups 6 to 11), respectively.
For data groups 6 to 11, the halves are equivalent to single pulses of the two peaks.
Therefore, we refer to them as the 'left' or the 'right' peak, while the sets of data groups 2 to 4 are referred to as the 'left' or 'right' half.
We note that we omit data group 5 as this is where the change in the profile occurred and thus the observations have neither clear a single peak nor a double peak in the folded profile.
\Cref{fig:SP_props_peaks} shows the width (top panel), mean flux density (middle panel), and fluence (bottom panel) distribution of the detected single pulses of the four new data sets.
The width distributions of the left half and the left peak are surprisingly similar since the overall width distribution in the time spans differ significantly, as shown in \cref{fig:SP_property_distributions}.
In both cases, there is a single peak distribution with a peak at around \SIrange{6}{9}{ms}, which is similar to the distributions of combined halves for the data rages 2 to 4 in \cref{fig:SP_property_distributions}.
Even the two distributions of the right half and the right peak are similar.
Both are close to the width distributions of data groups 6 to 11 (the double peaked observations).
The main difference between them is the increase in single pulses with a width less than \SI{1}{ms}, which is more prominent in the right peak.

For the mean flux density distributions, there is barely any difference between the two halves in the data groups 2 to 4, besides a few more faint pulses in the left half as a consequence of the width distribution.
Both are following the distributions of the respective time in \cref{fig:SP_property_distributions}.
For the data sets of the double peaked profiles (data groups 6 to 11), there is a significant difference between the two peaks. The single pulses in the right peak are distributed over a wider ranges of flux densities than the left half.
Comparing the left half and the left peak as well as the right half and the right peak, respectively, with each other, we can see that they are dominated by the time evolution (shown in \cref{fig:SP_property_distributions}) in contrast to the width distributions.

The fluence distributions are following those presented in \cref{fig:SP_property_distributions} for the respective data groups with only small differences between the left and the right half or peak.
For the data groups 2 to 4, 
the fluence of the left half is skewed to slightly higher fluences than the right half.
A similar difference is present for the peaks of the data groups 6 to 11, but at lower fluences.
In both cases, the difference is a consequence of the width distribution having a higher fraction of wide pulses, which results in more energetic single pulses of the left half.

\subsubsection{Phase dependence of the single pulse fluence}
\begin{figure}
\centering
\resizebox{\hsize}{!}{\includegraphics{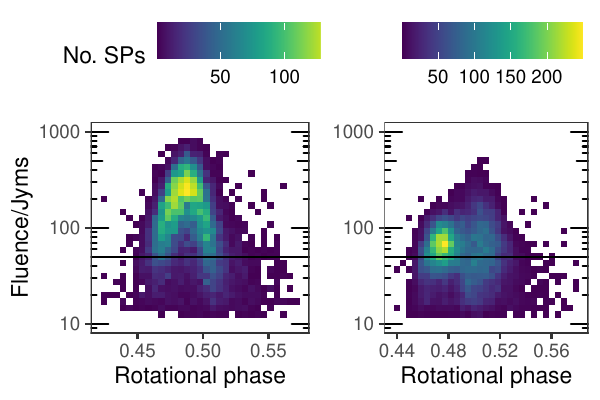}}
\caption{2D histogram of the fluence against the rotational phase for all single pulses detected in data group 2 (September 2020, single-peaked profile) and data group 6 (April 2021, double-peaked profile). The horizontal line corresponds to the completeness limit of a \SI{10}{ms} single pulse.}
\label{fig:SP_flux_vs_phase}
\end{figure}

While \cref{sec:clustering_SPs_rotation} presented the phase dependence of the single pulse width, here we look at the phase dependence of the single pulse fluence.
\Cref{fig:SP_flux_vs_phase} shows a two dimensional (2D) histogram of the single pulse fluence against rotational phase for data group 2 (left) and data group 6 (right).
For the single peaked profile in data group 2, the most energetic single pulses (i.e. those with the highest fluence) are found in the very centre of the profile window with fluences of about \SI{200}{Jy\,ms}.
At each phase, there is a preferred fluence, which is highest at the centre of the profile and decreases towards the edges.
The result is a 'banana shape' in the fluence phase distribution.
This holds true until the completeness limit is reached and our data set cannot represent the full single pulse population anymore.
It seems reasonable that the distribution continues into the incomplete area, and hence there may be many more single pulses, which we cannot detect as such at the profile edges.
These single pulses still contribute to the overall emission received from the magnetar when the time series is folded.
Thus, the emission at the outer parts of the profile from rotations without a detected single pulse in \cref{fig:mean_profiles} might be caused by these single pulses.
The 'banana shape' can be interpreted assuming that all single pulses are emitted within a beam that rotated through our line of sight (LOS).
The single pulses emitted off of the beam center appear weaker. 
Hence, the edges of the beam have lower fluence single pulses.

For the double-peaked data group 6, the highest amount of single pulses is found at 0.47 of the rotational phase with a fluence of \SI{60}{Jy\,ms}.
The second peak has single pulses with higher energies, but unlike the first peak, it shows a less concentrated distribution of single pulses with phase.
Moreover, the neither of peaks show a strong 'banana shape' distribution, which may be a consequence the incompleteness.
As for the single peaked case, this suggests that there are many more single pulses that we are not sensitive too at the outskirts of the emission window but also in the central regions as many pulses have been detected below the completeness limits in all phase angles.
Here, the interpretation is significantly hindered by the incompleteness.
With more sensitive instruments, it might be possible to detect a 'banana shape' for the individual components of the profile.

\section{Discussion}\label{sec:discussion}

\subsection{Integrating the parts - conclusions on the emission of XTE J1810-197} \label{sec:integrationg-the-parts}
In the previous section, we look at the emission of the magnetar XTE J1810-197 from several perspectives.
The aim of this section is to connect the individual results into an overall picture of the emission mechanism of the magnetar within phase 2.
The appearance of the folded profiles, as well as the single pulse properties, remain largely stable in consecutive observations with small variations on a observation to observation basis.
This is particularly the case for the pulse widths, which show a very smooth transition in their distribution as seen in \cref{fig:SP_property_distributions}.
Additionally, the fraction of the observed fluence from detected single pulses remains constant as seen in \cref{fig:energy_fraction}
All other emission properties, which are the profile shapes for the entire rotations (\cref{fig:mean_profiles}) and the single pulses (\cref{fig:SP_width_resolved_profiles}), as well as the fluence and mean flux density of the single pulses (\cref{fig:SP_property_distributions}) change abruptly around 17 March 2021 (related to the turn-off of the magnetar).

We can interpret these results by assuming that the emission originates from an emission region in the magnetosphere of the magnetar that co-rotates with the neutron star and the emission process can be modelled similarly to those of radio pulsars \citep[See for example][and references therein]{pulsarhandbook}.
Depending on how our LOS passes this emission region, different patches of the emission region become visible and the received flux changes: the emission appears brighter when the LOS passes the patch more edge on and appears fainter otherwise.
This is similar to the 'banana shape' when the emission region rotates in and out of our LOS seen in \cref{fig:SP_flux_vs_phase}.
If the emission region shifted slightly more away from our viewing angle during the turn off in mid March 2021, this would result in weaker single pulses as seen in \cref{fig:SP_flux_vs_phase} where the leading and trailing single pulses are significantly fainter.
This results in the changed fluence and mean flux density distributions seen in \cref{fig:SP_property_distributions}.
On the other hand, the width of the single pulses is not affected by this process, as this is related the size and how fast the emission region crosses our LOS.
The change in the width is caused by a different long-term process on the magnetar such as a long term-change in the emission height.
Since this shift also affects the overall received emission, the fraction of the total emission seen by detected single pulses remains constant.
Desvignes et all. (accepted) have shown that this magnetar undergoes precession and predict  that this changes how our LOS crosses the emission region and thus results in the different pulse profiles of the magnetar.
While this is a gradual change, a boundary in the magnetosphere might have been crossed a the turn off.
Another possible  process that lead to a shift the emission region is mode changing, which has been seen for normal radio pulsars for example by \cite{wang2007}.
\citet{rajwade2022} saw correlations between changes in the profile and nu-dot, which is characteristic of pulsar modeing \citet{lyne2010}, in Swift J1818.0-1607. Because we are unable to measure nu-dot in our data, we can only speculate that a similar process is occurring in XTE J1810-197. 

To further investigate the possibility of a shift of the emission region and its cause, the polarisation of the single pulses would be beneficial because this would allow for additional constraints to be placed on the emission region(s) for example from the polarisation position angle and its time and phase evolution.
This would also allow for further investigations of the long-term evolution of the precession observed by Desvignes et all. (accepted).
However, the current setup of the Stockert radio telescope does not enable polarisation information to be recorded.

As presented in \cref{fig:energy_fraction}, \SI{39}{\%} of the energy received from the magnetar during phase 2 cannot be linked to our observed single pulses.
The question remains what is the nature of the remaining emission, that is the profile from the rotations where no single pulse has been detected ('No SP' in \cref{fig:mean_profiles}).
There are several potential sources for single pulses that cannot be detected as such, but still show a contribution to the overall profile, as follows.
\begin{enumerate}
    \item Single pulses from the same population but below the detection threshold, for example, single pulse emission at the edges of the beam or from a weak additional component.
    The single pulse property distributions in \cref{fig:SP_property_distributions} suggest that these continue below the completeness limits.
    However, as the profiles without a detected single pulse deviate significantly from those with a detected single pulse, this can only account for some of the emission.
    \Cref{fig:SP_flux_vs_phase} shows that the fluence (and thus the S/N) of the pulses falls of at the edges of the beam.
    Hence, the broader profiles in the data groups 2 to 5 could be explained with this kind of emission, but it is unclear how strong these off-beam single pulses are for the profiles with two peaks.
    \item For very narrow single pulses:\ our sample does not include any pulses with a width less than \SI{0.65}{ms}. However, the width distributions in \cref{fig:SP_property_distributions} suggest that there are many narrow single pulses, as the distributions are cut by our completeness limits.
    These could be emitted in very limited rotational phases as the most narrow detectable single pulses in \cref{fig:SP_width_resolved_profiles}. For the data groups 6 to 11, where the 'No SP' profile peak in between the two peaks of the other profiles, the most narrow single pulses have been detected in between as well. Thus, adding more narrow pulses could explain the profile in this case. Similarly, the trailing part of the 'No SP' profile could be caused by narrow single pulses we cannot detect. Here, the narrow single pulses have been detected towards the end of duty cycle and the width distribution of the right half in \cref{fig:SP_props_peaks} suggests that the width distributions continues below the completeness limit.
    \item For very wide pulses (i.e. 'always on emission'), the decrease in single pulses wider than \SI{10}{ms} is real, as argued in \cref{sec:time_evo_properties}. Additionally, the extending the width distribution towards \SI{100}{ms} would mean that these pulses cover a large fraction of the duty cycle. This would be close to 'always on emission', which is a feature-less, faint emission over the full duty cycle.
    This kind of emission could also explain why the rotations without an detected single pulse exhibit fewer features than those with a detected single pulse.
\end{enumerate}
The case in points 1 and 2, which are testable on (archival) data of more sensitive telescopes. Point 2 would require a higher time resolution for these telescopes.
However, more sensitive telescopes would also find fainter emission in the average profile and, thus, they still cannot relate the entire emission to detected single pulses.
After recalculating the fraction of the overall emission that can be aligned to the detected single pulses, an upper limit for  the 'always on emission' case (as described in point 3) can be estimated.
For phase 1, we have detected only a few percent of the total emission with single pulses and no phase dependence of the pulse widths seems to be present.
Thus, we leave the source of the emission to the discussion where we take into account the observations of other telescopes.

\subsection{Relation to magnetars}
The radio emission of the magnetar XTE J1810-197 has been monitored by several authors in its previous (until 2008) and current radio outbursts.
The initial single pulses found by \citet{camilo2006} had widths of $\lesssim$ \SI{10}{ms} and peak flux densities of up to $\leq$ \SI{10}{Jy}.
Assuming a top hat pulse, this would give fluences of up to \SI{100}{Jy\,ms}, which is consistent with the highest energetic single pulses found in phase 1 of our data set but not with the phase 2, as these single pulses can reach significantly higher fluences.
The average flux density of the profile in this initial observations was at maximum \SI{10}{mJy} but mostly slightly less than \SI{1}{mJy}, which is significantly lower than in our data set.
The time evolution of the average flux density is similar to phase 1: the average flux density decreased after an initial high until it reached a stable low value.
Additionally, the pulse profiles presented in \citet{camilo2016} seem to show more structure and underwent major changes, where multiple components were visible at different times.
While we see variations from observation to observation, we have found a less dynamic change in the overall profile (besides the split in March 2021).

\citet{maan2019} studied the single pulses of XTE J1810-197 with the upgraded Giant Metrewave Radio Telescope (uGMRT) in December 2018 and February 2019. 
These authors found that the single pulses have peak flux densities of up to a few Jy and have rates of up to several thousand detected single pulses per hour.
This suggests that the about 1200 single pulses we found in phase 1 are the bright outliers of the single pulse population at that time. 
Additionally, \citet{maan2019} found that the single pulses align well with the overall average profile in phase.
Thus, their findings support our interpretation of the profiles we found in \cref{fig:mean_profiles} and \cref{fig:SP_width_resolved_profiles}.

\citet{caleb2022} observed XTE J1810-197 in both radio and X-ray for two years since its radio reappearance in December 2018 and thus overlap with our observations in the data groups 1 to 4.
A surprising result of the X-ray monitoring in their campaign is that the strong increase of the average radio flux density from May to September 2020 is lacking an increased activity in X-ray, but this would  show the lowest X-ray activity in campaign.
This phase has among the highest energetic single pulses and mean flux density detected over observational campaign as shown in \cref{fig:timeseries} and \cref{fig:SP_property_distributions}.
We inspected archival data from the Swift/XRT X-ray telescope in the rages August 2020 and February to April 2021 to see whether the second increase of radio loudness in February 2021 is related to an enhanced X-ray activity.
The X-ray activity of XTE J1810-197 is of about \SIrange{0.1}{0.15}{counts/s}, which is consistent with the cool down presented in \cite{borghese2021}. 
Therefore, the February increase is again lacking an corresponding increase in X-ray activity and the activity in the radio regime may evolve independently from the activity seen in X-ray, at least under the circumstances for XTE J1810-197.
This is in contrast to the commonly used search strategy for radio emission from magnetars, which targets magnetars after X-ray outbursts.
Thus, we argue that using X-ray as a trigger is a biased view on the potential of radio emission from magnetars and encourage regular independent monitoring of magnetars in the radio regime.

Additionally, \citet{caleb2022} claimed to have found giant pulses coming from this magnetar.
Their criterion for a giant pulse is for the average flux density of the on-pulse window to exceed ten times the average flux density, with the emission coming from a narrow phase range.
These giant pulses are dominated by a spiky emission that has a width of about \SI{10}{ms}.
It is important to notice that our definition of a single pulse differs to the one used by \citet{caleb2022} and this spiky emission is equivalent to our single pulse definition
The widths of the spiky emission fits well with the widths found in data group 2.
The observations where the giant pulses where found lay in the break between phase 1 and phase 2.
Hence, we cannot compare our findings directly to the giant pulse phase of the magnetar.
From the single pulse properties distributions in \cref{fig:SP_property_distributions}, we do not find a separate class of high fluence pulses.
Thus, the potential giant pulses might be the high energy end of the single pulse distribution.
Additionally, as the selection criterion is based on the average flux density of the entire on-pulse window, having additional weaker single pulses in the same rotation will give rise to a higher average flux density and would make the rotation match the criterion for giant pulses.

Generally, the width, fluence, and mean flux densities found for XTE J1810-197 in this work agree with the single pulse properties of other magnetars, for example \citet{wharton2019} for the galactic centre magnetar J1745-2900, \citet{levin2012} for J1622-4950,  \citet{esposito2020} and \citet{champion2020} for Swift J1818.0-1607.
While these works used more sensitive telescopes, the shorter and fewer observations limit the size of the single pulses data set for statistical analyses; namely,
they do detect significantly fainter single pulses, but the temporal change in the overall single pulse population cannot be studied, making it difficult to compare our findings with other works in this regard.

The magnetar SGR 1935+2154 became particularly relevant to the magnetar FRB connection after the detection of an FRB-like burst (FRB 20200428) by \citet{chimesgr19352020} and \citet{bochenek2020}.
In follow up observations, only a few more bursts have been detected by \citet{kirsten2021}, \citet{atel1935_chime} and \citet{atel1935_fast}.
The recent work by \citet{zhu2023} has found a larger amount of single pulses from this magnetar, which are orders of magnitude fainter and thus referred to as pulses rather than bursts.
The width of these single pulses are of the order of \SI{1}{ms}, which is significantly narrower than the single pulses detected in our work.
This could be caused by the shorter rotational period but also our incompleteness to pulses shorter than \SI{0.65}{ms}.
The SGR 1935+2154 single pulses fall in the RRAT regime of the transient phase space (\cref{fig:transient_phasespace}), which agrees with the single pulses we have found in this work.
Additionally, all single pulses from SGR 1935+2154 occur in a narrow phase window, which agrees with the observed behaviour of XTE J1810-197.
The FRB-like bursts from SGR 1935+2154, on the other hand, are not bound to the on-pulse window suggesting that the emission mechanism is different to the single pulses found by \citet{zhu2023}.
Hence, the single pulses from SGR 1935+2154 are analogous to the single pulses we found in our campaign.

\citet{kramer2023} propose that the periodicity of the micro structure duration ($P_\mu$) in radio transients is linearly related to their rotational periods.
The rotational period of XTE J1810-197 gives $P_\mu \approx$ \SI{5}{ms}.
From the examples shown in \cref{fig:SP_morphologies}, the periodicity of the pulses with a multi-burst morphology seems to be around this predicted values.
Thus, the relation  also holds at \SI{1.4}{GHz}, which is lower than the data used by \citet{kramer2023}, as those authors looked at XTE J1810-197 from \SIrange{4}{8}{GHz}.
However, an in-depth analysis especially regarding the time dependence of $P_\mu$ would be necessary to fully confirm the prediction.

\subsection{Relation to FRBs}
\begin{figure}
\centering
\resizebox{\hsize}{!}{\includegraphics{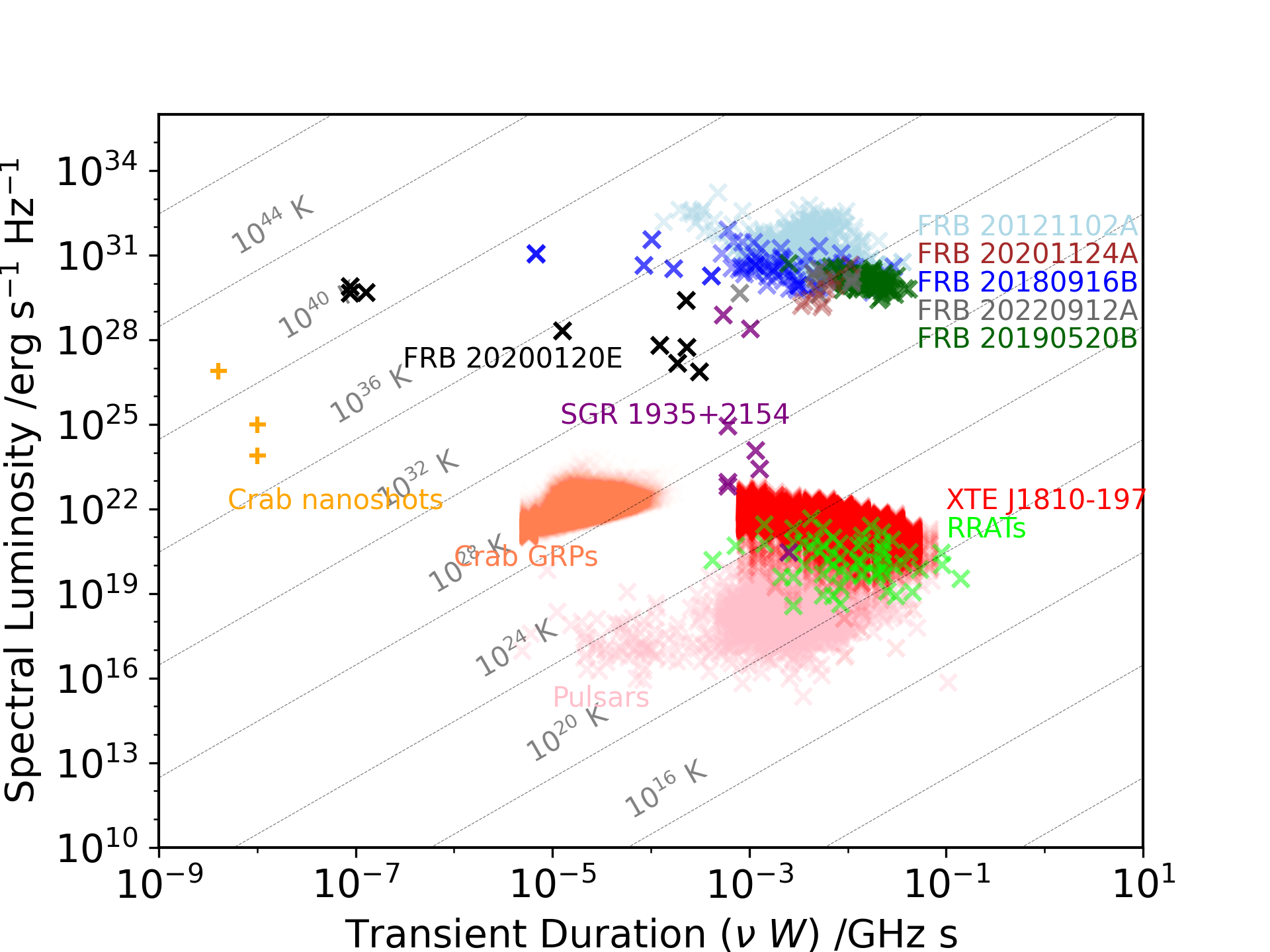}}
\caption{Transient phase space including the pulses from XTE J1810-197 presented in this work and some of the known repeating FRBs, RRAT pulses, and the bursts from SGR 1935+2154. Plot based on \citet{nimmo2022} and references their in with additional pulses from \citet{zhang2023}, \citet{niu2022} and \citet{hewitt2022}.}
\label{fig:transient_phasespace}
\end{figure}

The fluence and energy distributions of repeating FRBs are expected to follow a power law, but the observed single pulses do not seem to follow a power law.
Nevertheless, we can compare our observable to those seen in for FRBs.
To compare our single pulses to repeating FRB luminosities and durations, we display our single pulses from XTE J1810-197 (assuming a distance of \SI{2.5}{kpc} as reported by \citet{ding2020}), along with some well localised repeating FRBs and other radio transients (pulsars, Crab nanoshots, Crab Giant Radio Pulses, GRPs, and Rotating RAdio Transients, RRATs) in \cref{fig:transient_phasespace}, which is based on \citet{nimmo2022}\footnote{\url{https://github.com/KenzieNimmo/tps}}.
Our brightest pulses are about \SI{e22}{erg/s/Hz} and, hence, six orders of magnitude less bright than FRB 20200428 (from SGR 1935+2154).
In comparison to the other (extragalactic) FRBs, the luminosity is about four orders of magnitude below the bursts seen from the repeater in M81 (FRB20200120E) and about seven orders of magnitude below the parameters space covered by most other repeaters.

The extremely bright burst from SGR 1935+2154 has been a rare event so far.
Furthermore, the distinction between pulses and bursts proposed by \citet{zhu2023} indicates that bursts only occur under specific conditions.
It is unclear what conditions are required for such events and whether XTE 1810-197 can satisfy them.
Therefore, the duration of our observation campaign may not have been long enough to capture extremely rare, but very bright events.

\begin{figure}
\centering
\resizebox{\hsize}{!}{\includegraphics{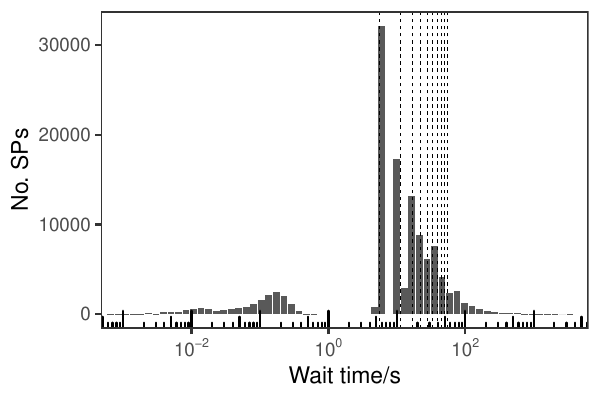}}
\caption{Waiting time distribution for the detected single pulses of the observations since September 2020. The vertical lines mark multiples of the rotational period (\SI{5.542}{s}).}
\label{fig:waiting_time}
\end{figure}

Additionally, we have not found any of the characteristic spectral features of the known repeating FRBs, such as band-limited emission with downward frequency drifts. 
However, we see similar morphologies in the time series, for example, from FRB 20121102A by \citet{jahns2023} or in the CHIME/FRB sample (\citet{pleunis2021}). 
The waiting time distribution presented in \citet{jahns2023} has similar features as the waiting time distribution of our detected single pulses, which is shown in \cref{fig:waiting_time}.
For our observations, there are pulses with waiting times from \SIrange{2}{400}{ms}(i.e. less than the rotational period) and single pulses with waiting times larger than \SI{5}{s}, with peaks at integer multiples of the rotational period (indicated by the vertical lines in \cref{fig:waiting_time}).
The waiting times within a single rotation split into two peaks at about \SI{25}{ms} and \SI{200}{ms}.
\citet{jahns2023} find bursts from FRB20121102A in three peaks.
Two of them are in the range of \SIrange{1}{100}{ms}, where the Poisson interpretation predicts zero bursts, and the third in the range \SIrange{1}{100}{s} with a few bursts in between.
Hence, the waiting time distribution is continuous in contrast to our observations, which show a clear gap.
The peaks at \SIrange{1}{100}{ms} can be split into sub-bursts (median \SI{4}{ms}) and separate bursts (median \SI{22}{ms}.
Thus, the waiting time distribution consists of three peaks as for XTE J1810-197.
The first two peaks are each about an order of magnitude shorter for the FRB than for the magnetar.
In both cases, the first peak can be linked to sub-bursts and sub-pulses and is thus a characteristic of the emission process.
The different time scales might indicate different environments in which the pulse or burst is produced or a difference in the duration of the rotational period.
\citet{jahns2023} argued that the second peak (indicating the duty cycle for the magnetar in our case) is related to the physical process that produces the bursts that is has a similar interpretation as the sub-bursts.
The third peak for the magnetar is a consequence of its clear periodicity, but for the FRB, it is interpreted as a consequence of the Poissonian nature of the emission.
Therefore, the presence of detected FRBs between the emission scale peak and the Poisson peak is particularly challenging for a rotational progenitor model.
A potential explanation could be a very wide duty cycle or the possibility that FRBs are not bound to the emission window, as proposed by \citet{zhu2023}.


\section{Conclusions} \label{sec:conclusion}

We report the long-term monitoring of the magnetar XTE J1810-197 with the Stockert radio telescope between December 2018 and November 2021. Here, we have found 115,000 single pulses using our presented filtering method.
During the observational campaign, the magnetar shows different emission properties, most notably between the 2018/19 observations (phase 1) and the 2020/21 observations (phase 2).
\begin{enumerate}
    \item Phase 1 and phase 2 observations showed folded profiles with comparable flux densities, while the rate of detected single pulses differ significantly with a few per hour in phase 1 and hundreds per hour in phase 2.
    \item In phase 2, the emission properties show two distinct groups split by short phase of inactivity around mid March 2021.
Up to March 2021, the profile consists of a single visible peak and wide (single mode distribution peaking around \SI{10}{ms}) single pulses with fluences of up to \SI{1000}{Jy\,ms}, while the profiles after mid March 2021 show two distinct peaks and the single pulses are more narrow and less energetic.
    \item Based on the similarities between the single pulse width distribution of the left half and the left peak and right half and right peak, respectively, we argue that this could be caused by a drift of the emission regions in the magnetosphere.
This drift caused the profile to split as well as lower fluences and mean flux densities, while the widths do not change abruptly and follow a long-term trend.
A study of the polarisation properties of the single pulses would be a test of this hypothesis.
    \item We can link about \SI{61}{\%} of the received emission in phase 2 to rotations to detected single pulses, which have very similar profiles to the single pulses emission itself.
The remaining emission can be due to single pulses we cannot detect (extreme widths or faint) but we cannot entirely rule out an 'always on emission' case.
    \item The emission at the edges of the rotations without a detected single pulse can be explained by the phase dependence of the fluence of the single pulses as the beam passes our LOS.
    \item The fluence distributions of the single pulses do not follow a power law distribution, but they are otherwise similar to the single pulses seen for other magnetars.
    \item The radio outbursts of the XTE J1810-197 in the September 2020 and the February 2021 are not associated with a X-ray activity, indicating that radio is not always following X-ray activity.
\end{enumerate}
We will continue to monitor the magnetar XTE J1810-197 with the Stockert radio telescope on a regular basis and encourage observations with telescopes of higher sensitivity as well as polarisation capabilities.
This will yield a better understanding of the emission mechanism of the magnetar, which could potentially help to improve the overall understanding of other neutron star-related radio emission, as well as that of FRBs.

 \section*{Data availability\footnote{Other data products and the raw data are stored in the archive and can be shared upon reasonable request by the authors.}}
 The pulse rate and average flux density per observation (Table A1) and a list of single pulses with the properties discussed in this work (Table A2) are only available online at the CDS via anonymous ftp to cdsarc.u-strasbg.fr (130.79.128.5) or via http://cdsweb.u-strasbg.fr/cgi-bin/qcat?J/A+A/ and at the Max Planck digital library\footnote{\url{https://doi.org/10.17617/3.7WQANS}}.
 Table A1 contains the following information. Column 1 gives the date of the observation, column 2 gives the MJD of the start of the observation, column 3 gives the number of pulses detected by the filtering, column 4 gives the uncertainty for the detected number of pulses, column 5 (Jy) gives the mean flux density of the folded profile and column 6 (Jy) gives the uncertainty for the mean flux density.
 Table A2 contains the following information. Column 1 gives the MJD of the start of the observation, column 2 gives the time (in s) since the start of the observations when the single pulse was detected, column 3 gives the phase of the centre of the pulse (in s), column 4 gives the mean flux density of the pulse (Jy), column 5 gives the fluence of the single pulse (Jyms), column 6 gives the width of the single pulse (ms), and column 7 indicates whether the pulse was detected in the left or right half/peak (1=left, 2=right).

\begin{acknowledgements}
      The authors acknowledge the many hours of operating provided by the members of Astropeiler Stockert e.V. on a purely voluntary bases. We also acknowledge the support of the Stockert telescope and the funding of its refurbishment by the Nordrhein-Westfalen Stiftung.
      LGS is a Lise Meitner Max Planck research group leader and acknowledges support from the Max Planck Society.
      The authors thankfully acknowledge the provided ephemerides from the Jodrell bank observatory team by Andrew Lyne, Manisha Caleb, Kaustubh Rajwade and Benjamin Stappers.
      This research has made use of data, software and/or web tools obtained from the High Energy Astrophysics Science Archive Research Center (HEASARC), a service of the Astrophysics Science Division at NASA/GSFC and of the Smithsonian Astrophysical Observatory's High Energy Astrophysics Division.
\end{acknowledgements}

   \bibliographystyle{aa} 
   \bibliography{refs.bib} 

\begin{thebibliography}{43}
\expandafter\ifx\csname natexlab\endcsname\relax\def\natexlab#1{#1}\fi

\bibitem[{Agarwal {et~al.}(2020)Agarwal, Aggarwal, Burke-Spolaor, Lorimer, \&
  Garver-Daniels}]{agarwal2020}
Agarwal, D., Aggarwal, K., Burke-Spolaor, S., Lorimer, D.~R., \&
  Garver-Daniels, N. 2020, Monthly Notices of the Royal Astronomical Society,
  497, 1661

\bibitem[{{Baars} {et~al.}(1977){Baars}, {Genzel}, {Pauliny-Toth}, \&
  {Witzel}}]{baars1977}
{Baars}, J.~W.~M., {Genzel}, R., {Pauliny-Toth}, I.~I.~K., \& {Witzel}, A.
  1977, \aap, 61, 99

\bibitem[{{Barr} {et~al.}(2013){Barr}, {Champion}, {Kramer}, {Eatough},
  {Freire}, {Karuppusamy}, {Lee}, {Verbiest}, {Bassa}, {Lyne}, {Stappers},
  {Lorimer}, \& {Klein}}]{barr2013}
{Barr}, E.~D., {Champion}, D.~J., {Kramer}, M., {et~al.} 2013, \mnras, 435,
  2234

\bibitem[{{Bochenek} {et~al.}(2020){Bochenek}, {Ravi}, {Belov}, {Hallinan},
  {Kocz}, {Kulkarni}, \& {McKenna}}]{bochenek2020}
{Bochenek}, C.~D., {Ravi}, V., {Belov}, K.~V., {et~al.} 2020, \nat, 587, 59

\bibitem[{{Borghese} {et~al.}(2021){Borghese}, {Rea}, {Turolla}, {Rigoselli},
  {Alford}, {Gotthelf}, {Burgay}, {Possenti}, {Zane}, {Coti Zelati}, {Perna},
  {Esposito}, {Mereghetti}, {Vigan{\`o}}, {Tiengo}, {G{\"o}tz}, {Ibrahim},
  {Israel}, {Pons}, \& {Sathyaprakash}}]{borghese2021}
{Borghese}, A., {Rea}, N., {Turolla}, R., {et~al.} 2021, \mnras, 504, 5244

\bibitem[{{Caleb} {et~al.}(2022){Caleb}, {Rajwade}, {Desvignes}, {Stappers},
  {Lyne}, {Weltevrede}, {Kramer}, {Levin}, \& {Surnis}}]{caleb2022}
{Caleb}, M., {Rajwade}, K., {Desvignes}, G., {et~al.} 2022, \mnras, 510, 1996

\bibitem[{{Camilo} {et~al.}(2016){Camilo}, {Ransom}, {Halpern}, {Alford},
  {Cognard}, {Reynolds}, {Johnston}, {Sarkissian}, \& {van
  Straten}}]{camilo2016}
{Camilo}, F., {Ransom}, S.~M., {Halpern}, J.~P., {et~al.} 2016, 820, 110

\bibitem[{{Camilo} {et~al.}(2006){Camilo}, {Ransom}, {Halpern}, {Reynolds},
  {Helfand}, {Zimmerman}, \& {Sarkissian}}]{camilo2006}
{Camilo}, F., {Ransom}, S.~M., {Halpern}, J.~P., {et~al.} 2006, 442, 892

\bibitem[{{Champion} {et~al.}(2020){Champion}, {Cognard}, {Cruces},
  {Desvignes}, {Jankowski}, {Karuppusamy}, {Keith}, {Kouveliotou}, {Kramer},
  {Liu}, {Lyne}, {Mickaliger}, {O'Connor}, {Parthasarathy}, {Porayko},
  {Rajwade}, {Stappers}, {Torne}, {van der Horst}, \&
  {Weltevrede}}]{champion2020}
{Champion}, D., {Cognard}, I., {Cruces}, M., {et~al.} 2020, \mnras, 498, 6044

\bibitem[{{CHIME/FRB Collaboration} {et~al.}(2020){CHIME/FRB Collaboration},
  {Andersen}, {Bandura}, {Bhardwaj}, {Bij}, {Boyce}, {Boyle}, {Brar},
  {Cassanelli}, {Chawla}, {Chen}, {Cliche}, {Cook}, {Cubranic}, {Curtin},
  {Denman}, {Dobbs}, {Dong}, {Fandino}, {Fonseca}, {Gaensler}, {Giri}, {Good},
  {Halpern}, {Hill}, {Hinshaw}, {H{\"o}fer}, {Josephy}, {Kania}, {Kaspi},
  {Landecker}, {Leung}, {Li}, {Lin}, {Masui}, {McKinven}, {Mena-Parra},
  {Merryfield}, {Meyers}, {Michilli}, {Milutinovic}, {Mirhosseini},
  {M{\"u}nchmeyer}, {Naidu}, {Newburgh}, {Ng}, {Patel}, {Pen},
  {Pinsonneault-Marotte}, {Pleunis}, {Quine}, {Rafiei-Ravandi}, {Rahman},
  {Ransom}, {Renard}, {Sanghavi}, {Scholz}, {Shaw}, {Shin}, {Siegel}, {Singh},
  {Smegal}, {Smith}, {Stairs}, {Tan}, {Tendulkar}, {Tretyakov}, {Vanderlinde},
  {Wang}, {Wulf}, \& {Zwaniga}}]{chimesgr19352020}
{CHIME/FRB Collaboration}, {Andersen}, B.~C., {Bandura}, K.~M., {et~al.} 2020,
  \nat, 587, 54

\bibitem[{{Ding} {et~al.}(2020){Ding}, {Deller}, {Lower}, {Flynn},
  {Chatterjee}, {Brisken}, {Hurley-Walker}, {Camilo}, {Sarkissian}, \&
  {Gupta}}]{ding2020}
{Ding}, H., {Deller}, A.~T., {Lower}, M.~E., {et~al.} 2020, \mnras, 498, 3736

\bibitem[{{Duncan} \& {Thompson}(1992)}]{duncan1992}
{Duncan}, R.~C. \& {Thompson}, C. 1992, \apjl, 392, L9

\bibitem[{{Esposito} {et~al.}(2020){Esposito}, {Rea}, {Borghese}, {Coti
  Zelati}, {Vigan{\`o}}, {Israel}, {Tiengo}, {Ridolfi}, {Possenti}, {Burgay},
  {G{\"o}tz}, {Pintore}, {Stella}, {Dehman}, {Ronchi}, {Campana},
  {Garcia-Garcia}, {Graber}, {Mereghetti}, {Perna}, {Rodr{\'\i}guez Castillo},
  {Turolla}, \& {Zane}}]{esposito2020}
{Esposito}, P., {Rea}, N., {Borghese}, A., {et~al.} 2020, \apjl, 896, L30

\bibitem[{{Halpern} {et~al.}(2005){Halpern}, {Gotthelf}, {Becker}, {Helfand},
  \& {White}}]{halpern2005}
{Halpern}, J.~P., {Gotthelf}, E.~V., {Becker}, R.~H., {Helfand}, D.~J., \&
  {White}, R.~L. 2005, \apjl, 632, L29

\bibitem[{{Hessels} {et~al.}(2019){Hessels}, {Spitler}, {Seymour}, {Cordes},
  {Michilli}, {Lynch}, {Gourdji}, {Archibald}, {Bassa}, {Bower}, {Chatterjee},
  {Connor}, {Crawford}, {Deneva}, {Gajjar}, {Kaspi}, {Keimpema}, {Law},
  {Marcote}, {McLaughlin}, {Paragi}, {Petroff}, {Ransom}, {Scholz}, {Stappers},
  \& {Tendulkar}}]{hessels2019}
{Hessels}, J.~W.~T., {Spitler}, L.~G., {Seymour}, A.~D., {et~al.} 2019, \apjl,
  876, L23

\bibitem[{Hewitt {et~al.}(2022)Hewitt, Snelders, Hessels, Nimmo, Jahns,
  Spitler, Gourdji, Hilmarsson, Michilli, Ould-Boukattine, Scholz, \&
  Seymour}]{hewitt2022}
Hewitt, D.~M., Snelders, M.~P., Hessels, J. W.~T., {et~al.} 2022, Monthly
  Notices of the Royal Astronomical Society, 515, 3577

\bibitem[{{Ibrahim} {et~al.}(2004){Ibrahim}, {Markwardt}, {Swank}, {Ransom},
  {Roberts}, {Kaspi}, {Woods}, {Safi-Harb}, {Balman}, {Parke}, {Kouveliotou},
  {Hurley}, \& {Cline}}]{ibrahim2004}
{Ibrahim}, A.~I., {Markwardt}, C.~B., {Swank}, J.~H., {et~al.} 2004, 609, L21

\bibitem[{{Jahns} {et~al.}(2023){Jahns}, {Spitler}, {Nimmo}, {Hewitt},
  {Snelders}, {Seymour}, {Hessels}, {Gourdji}, {Michilli}, \&
  {Hilmarsson}}]{jahns2023}
{Jahns}, J.~N., {Spitler}, L.~G., {Nimmo}, K., {et~al.} 2023, \mnras, 519, 666

\bibitem[{{Kirsten} {et~al.}(2021){Kirsten}, {Snelders}, {Jenkins}, {Nimmo},
  {van den Eijnden}, {Hessels}, {Gawro{\'n}ski}, \& {Yang}}]{kirsten2021}
{Kirsten}, F., {Snelders}, M.~P., {Jenkins}, M., {et~al.} 2021, Nature
  Astronomy, 5, 414

\bibitem[{Kramer {et~al.}(2023)Kramer, Liu, Desvignes, Karuppusamy, \&
  Stappers}]{kramer2023}
Kramer, M., Liu, K., Desvignes, G., Karuppusamy, R., \& Stappers, B.~W. 2023,
  Nature Astronomy

\bibitem[{{Levin} {et~al.}(2012){Levin}, {Bailes}, {Bates}, {Bhat}, {Burgay},
  {Burke-Spolaor}, {D'Amico}, {Johnston}, {Keith}, {Kramer}, {Milia},
  {Possenti}, {Stappers}, \& {van Straten}}]{levin2012}
{Levin}, L., {Bailes}, M., {Bates}, S.~D., {et~al.} 2012, \mnras, 422, 2489

\bibitem[{{Levin} {et~al.}(2019){Levin}, {Lyne}, {Desvignes}, {Eatough},
  {Karuppusamy}, {Kramer}, {Mickaliger}, {Stappers}, \&
  {Weltevrede}}]{levin2019}
{Levin}, L., {Lyne}, A.~G., {Desvignes}, G., {et~al.} 2019, \mnras, 488, 5251

\bibitem[{{Lorimer}(2011)}]{lorimer2011}
{Lorimer}, D.~R. 2011, {SIGPROC: Pulsar Signal Processing Programs},
  Astrophysics Source Code Library, record ascl:1107.016

\bibitem[{Lorimer {et~al.}(2007)Lorimer, Bailes, McLaughlin, Narkevic, \&
  Crawford}]{lorimer2007}
Lorimer, D.~R., Bailes, M., McLaughlin, M.~A., Narkevic, D.~J., \& Crawford, F.
  2007, Science, 318, 777–780

\bibitem[{{Lorimer} \& {Kramer}(2004)}]{pulsarhandbook}
{Lorimer}, D.~R. \& {Kramer}, M. 2004, {Handbook of Pulsar Astronomy}, Vol.~4

\bibitem[{{Lower} {et~al.}(2020){Lower}, {Gupta}, {Flynn}, {Bailes}, {Jameson},
  {Farah}, {Bateman}, {Campbell-Wilson}, {Day}, {Deller}, {Green}, {Mandlik},
  {Oslowski}, {Parthasarathy}, {Price}, {Sutherland}, {Temby}, {Torr},
  {Urquhart}, {Krishnan}, \& {Venville}}]{ATel13840}
{Lower}, M.~E., {Gupta}, V., {Flynn}, C., {et~al.} 2020, The Astronomer's
  Telegram, 13840, 1

\bibitem[{{Lyne} {et~al.}(2010){Lyne}, {Hobbs}, {Kramer}, {Stairs}, \&
  {Stappers}}]{lyne2010}
{Lyne}, A., {Hobbs}, G., {Kramer}, M., {Stairs}, I., \& {Stappers}, B. 2010,
  Science, 329, 408

\bibitem[{{Lyne} {et~al.}(2018){Lyne}, {Levin}, {Stappers}, {Mickaliger},
  {Desvignes}, \& {Kramer}}]{lyne2018}
{Lyne}, A., {Levin}, L., {Stappers}, B., {et~al.} 2018, The Astronomer's
  Telegram, 12284, 1

\bibitem[{{Maan} {et~al.}(2019){Maan}, {Joshi}, {Surnis}, {Bagchi}, \&
  {Manoharan}}]{maan2019}
{Maan}, Y., {Joshi}, B.~C., {Surnis}, M.~P., {Bagchi}, M., \& {Manoharan},
  P.~K. 2019, \apjl, 882, L9

\bibitem[{{Nimmo} {et~al.}(2022){Nimmo}, {Hessels}, {Kirsten}, {Keimpema},
  {Cordes}, {Snelders}, {Hewitt}, {Karuppusamy}, {Archibald}, {Bezrukovs},
  {Bhardwaj}, {Blaauw}, {Buttaccio}, {Cassanelli}, {Conway}, {Corongiu},
  {Feiler}, {Fonseca}, {Forss{\'e}n}, {Gawro{\'n}ski}, {Giroletti}, {Kharinov},
  {Leung}, {Lindqvist}, {Maccaferri}, {Marcote}, {Masui}, {Mckinven},
  {Melnikov}, {Michilli}, {Mikhailov}, {Ng}, {Orbidans}, {Ould-Boukattine},
  {Paragi}, {Pearlman}, {Petroff}, {Rahman}, {Scholz}, {Shin}, {Smith},
  {Stairs}, {Surcis}, {Tendulkar}, {Vlemmings}, {Wang}, {Yang}, \&
  {Yuan}}]{nimmo2022}
{Nimmo}, K., {Hessels}, J.~W.~T., {Kirsten}, F., {et~al.} 2022, Nature
  Astronomy, 6, 393

\bibitem[{{Niu} {et~al.}(2022){Niu}, {Aggarwal}, {Li}, {Zhang}, {Chatterjee},
  {Tsai}, {Yu}, {Law}, {Burke-Spolaor}, {Cordes}, {Zhang}, {Ocker}, {Yao},
  {Wang}, {Feng}, {Niino}, {Bochenek}, {Cruces}, {Connor}, {Jiang}, {Dai},
  {Luo}, {Li}, {Miao}, {Niu}, {Anna-Thomas}, {Sydnor}, {Stern}, {Wang}, {Yuan},
  {Yue}, {Zhou}, {Yan}, {Zhu}, \& {Zhang}}]{niu2022}
{Niu}, C.~H., {Aggarwal}, K., {Li}, D., {et~al.} 2022, \nat, 606, 873

\bibitem[{{Olausen} \& {Kaspi}(2014)}]{olausen2014}
{Olausen}, S.~A. \& {Kaspi}, V.~M. 2014, \apjs, 212, 6

\bibitem[{Platts {et~al.}(2019)Platts, Weltman, Walters, Tendulkar, Gordin, \&
  Kandhai}]{platts2019}
Platts, E., Weltman, A., Walters, A., {et~al.} 2019, Physics Reports, 821, 1

\bibitem[{{Pleunis} \& {CHIME/FRB Collaboration}(2020)}]{atel1935_chime}
{Pleunis}, Z. \& {CHIME/FRB Collaboration}. 2020, The Astronomer's Telegram,
  14080, 1

\bibitem[{Pleunis {et~al.}(2021)Pleunis, Good, Kaspi, Mckinven, Ransom, Scholz,
  Bandura, Bhardwaj, Boyle, Brar, Cassanelli, Chawla, Fengqiu, Dong, Fonseca,
  Gaensler, Josephy, Kaczmarek, Leung, Lin, Masui, Mena-Parra, Michilli, Ng,
  Patel, Rafiei-Ravandi, Rahman, Sanghavi, Shin, Smith, Stairs, \&
  Tendulkar}]{pleunis2021}
Pleunis, Z., Good, D.~C., Kaspi, V.~M., {et~al.} 2021, Fast Radio Burst
  Morphology in the First CHIME/FRB Catalog

\bibitem[{Rajwade {et~al.}(2022)Rajwade, Stappers, Lyne, Shaw, Mickaliger, Liu,
  Kramer, Desvignes, Karuppusamy, Enoto, Güver, Hu, \& Surnis}]{rajwade2022}
Rajwade, K.~M., Stappers, B.~W., Lyne, A.~G., {et~al.} 2022, Monthly Notices of
  the Royal Astronomical Society, 512, 1687

\bibitem[{{Rea} {et~al.}(2012){Rea}, {Pons}, {Torres}, \& {Turolla}}]{rea2012}
{Rea}, N., {Pons}, J.~A., {Torres}, D.~F., \& {Turolla}, R. 2012, \apjl, 748,
  L12

\bibitem[{{Spitler} {et~al.}(2012){Spitler}, {Cordes}, {Chatterjee}, \&
  {Stone}}]{spitler2012}
{Spitler}, L.~G., {Cordes}, J.~M., {Chatterjee}, S., \& {Stone}, J. 2012, \apj,
  748, 73

\bibitem[{Wang {et~al.}(2007)Wang, Manchester, \& Johnston}]{wang2007}
Wang, N., Manchester, R.~N., \& Johnston, S. 2007, Monthly Notices of the Royal
  Astronomical Society, 377, 1383

\bibitem[{Wharton {et~al.}(2019)Wharton, Chatterjee, Cordes, Bower, Butler,
  Deller, Demorest, Lazio, \& Ransom}]{wharton2019}
Wharton, R.~S., Chatterjee, S., Cordes, J.~M., {et~al.} 2019, The Astrophysical
  Journal, 875, 143

\bibitem[{{Zhang} {et~al.}(2020){Zhang}, {Jiang}, {Men}, {Wang}, {Xu}, {Xu},
  {Niu}, {Zhou}, {Guan}, {Han}, {Jiang}, {Lee}, {Li}, {Lin}, {Niu}, {Wang},
  {Wang}, {Xu}, {Yu}, {Zhang}, \& {Zhu}}]{atel1935_fast}
{Zhang}, C.~F., {Jiang}, J.~C., {Men}, Y.~P., {et~al.} 2020, The Astronomer's
  Telegram, 13699, 1

\bibitem[{Zhang {et~al.}(2023)Zhang, Wang, Yang, Li, Geng, Tang, Chang, Luo,
  Wang, Wu, Dai, \& Zhang}]{zhang2023}
Zhang, S.~B., Wang, J.~S., Yang, X., {et~al.} 2023, A bright radio burst from
  FRB 20200120E in a globular cluster of the nearby galaxy M81

\bibitem[{Zhu {et~al.}(2023)Zhu, Xu, Zhou, Lin, Wang, Wang, Zhang, Niu, Chen,
  Li, Meng, Lee, Zhang, Feng, Ge, Gö{\u{g} }ü{\c{s}}, Guan, Han, Jiang,
  Jiang, Kouveliotou, Li, Miao, Miao, Men, Niu, Wang, Wang, Xu, Xu, Xue, Yang,
  Yu, Yuan, Yue, Zhang, \& Zhang}]{zhu2023}
Zhu, W., Xu, H., Zhou, D., {et~al.} 2023, Science Advances, 9

\end{thebibliography}

\begin{appendix}
    \section{Single pulse morphologies}
\begin{figure}
    \centering
    \includegraphics[width=17cm]{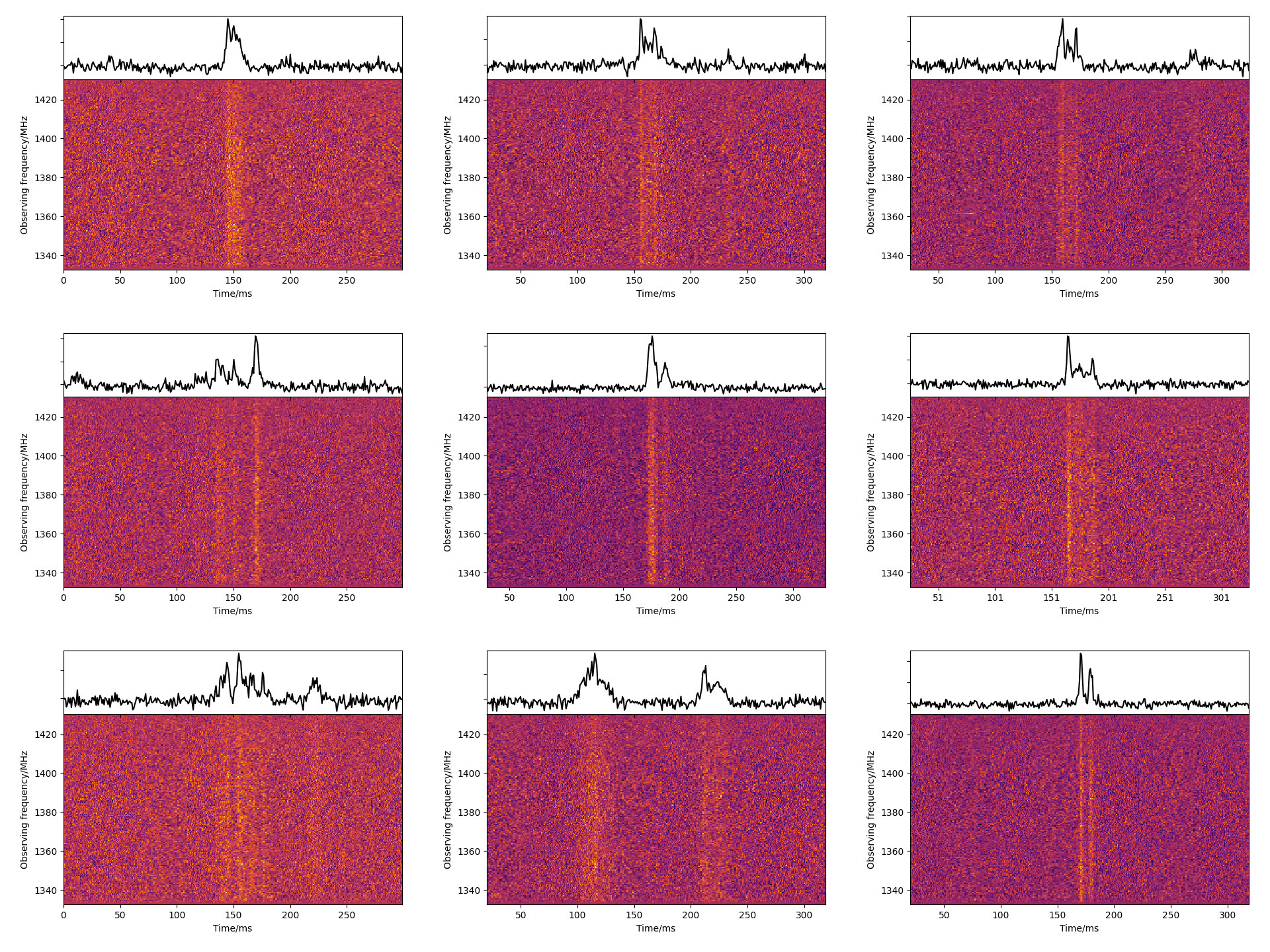}
    \caption{Examples of the different morphologies in the dynamic spectrum and the time series of the detected single pulses from the observation on 29 September 2020.}
    \label{fig:SP_morphologies}
\end{figure}

\Cref{fig:SP_morphologies} shows examples of different morphologies of single pulses all seen within a single observation on 29 September 2020.
\end{appendix}

\end{document}